\documentclass[twocolumn,aps,prd,groupedaddress,superscriptaddress,amsmath,amssymb,noeprint,nofootinbib,10pt]{revtex4-2}
\usepackage[T1]{fontenc}
\usepackage{graphicx}
\graphicspath{{./Figures/}}
\usepackage{dcolumn}
\usepackage{bm}
\usepackage[dvipsnames]{xcolor}
\usepackage{hyperref}
\hypersetup{colorlinks,linkcolor={MidnightBlue},citecolor={MidnightBlue},urlcolor={MidnightBlue}}  

\usepackage[table,xcdraw]{xcolor}
\usepackage{tabularray}

\usepackage{float}
\usepackage{wrapfig}

\usepackage{tensor}

\usepackage{physics}
\usepackage{amsfonts}
\usepackage{verbatim}
\usepackage{amsmath}
\usepackage{csquotes}
\usepackage{color}
\usepackage{soul}
\usepackage{amsthm}
\usepackage{bm}
\usepackage{graphicx}
\usepackage[normalem]{ulem}
\usepackage{mathtools}

\usepackage{yfonts}

\usepackage{verbatim}

\newenvironment{equations}
{\begin{equation}\begin{aligned}}
		{\end{aligned}\end{equation}\ignorespacesafterend}

\newenvironment{equations*}
{\begin{equation*}\begin{aligned}}
		{\end{aligned}\end{equation*}\ignorespacesafterend}

\newcommand{\prt}[1]{\left(#1\right)}

\newcommand{\Schr}{Schr\"{o}dinger\ }

\newcommand{\dg}{\dagger}
\newcommand{\tl}{\tilde}

\newcommand{\td}{\tilde{d}}

\newcommand{\prtq}[1]{\left[#1\right]}

\newcommand{\prtg}[1]{\left\{#1\right\}}

\newcommand{\prtqB}[1]{\Bigg[#1\Bigg]}

\newcommand{\sign}{\mathrm{sign}}

\newcommand{\ghm}{\textgoth{m}}

\newcommand{\mcL}{\mathcal{L}}

\newcommand{\mcV}{\mathcal{V}}

\newcommand{\mcD}{\mathcal{D}}

\newcommand{\mcT}{\mathcal{T}}

\newcommand{\mcF}{\mathcal{F}}

\newcommand{\mbE}{\mathbb{E}}

\newcommand{\hp}{\hat{p}}
\newcommand{\hq}{\hat{q}}

\newcommand{\hmu}{\hat{\mu}}

\newcommand{\tmuR}{\tilde{\mu}_{r_C}}

\newcommand{\hA}{\hat{A}}

\newcommand{\hF}{\hat{F}}

\newcommand{\hV}{\hat{V}}

\newcommand{\hL}{\hat{L}}

\newcommand{\bx}{\mathbf{x}}
\newcommand{\by}{\mathbf{y}}
\newcommand{\bz}{\mathbf{z}}

\newcommand{\bk}{\mathbf{k}}

\newcommand{\bd}{\mathbf{d}}

\newcommand{\bX}{\mathbf{X}}
\newcommand{\bY}{\mathbf{Y}}

\newcommand{\bD}{\mathbf{D}}

\newcommand{\bv}{\mathbf{v}}

\newcommand{\hbq}{\hat{\mathbf{q}}}

\newcommand{\hbP}{\hat{\mathbf{P}}}
\newcommand{\hbQ}{\hat{\mathbf{Q}}}

\makeatletter 

\renewcommand\onecolumngrid{
\do@columngrid{one}{\@ne}%
\def\set@footnotewidth{\onecolumngrid}
\def\footnoterule{\kern-6pt\hrule width 1.5in\kern6pt}%
}

\renewcommand\twocolumngrid{
\def\footnoterule{
\dimen@\skip\footins\divide\dimen@\thr@@
\kern-\dimen@\hrule width.5in\kern\dimen@}
\do@columngrid{mlt}{\tw@}
}%

\makeatother    

\begin{document}

\title{Hybrid Classical-Quantum Newtonian Gravity with Stable Vacuum}

\author{Nicol\`{o} Piccione}
\email{nicolo'.piccione@units.it}
\affiliation{Department of Physics, University of Trieste, Strada Costiera 11, 34151 Trieste, Italy}
\affiliation{Istituto Nazionale di Fisica Nucleare, Trieste Section, Via Valerio 2, 34127 Trieste, Italy}

\author{Angelo Bassi}
\affiliation{Department of Physics, University of Trieste, Strada Costiera 11, 34151 Trieste, Italy}
\affiliation{Istituto Nazionale di Fisica Nucleare, Trieste Section, Via Valerio 2, 34127 Trieste, Italy}

\begin{abstract}
We investigate the Gravitational Poissonian Spontaneous Localization (GPSL) model, a hybrid classical-quantum model in which classical Newtonian gravity emerges from stochastic collapses of the mass density operator, and consistently couples to quantum matter. Unlike models based on continuous weak measurement schemes, we show that GPSL ensures vacuum stability; this, together with its applicability to identical particles and fields, makes it a promising candidate for a relativistic generalization. 
We analyze the model’s general properties, and compare its predictions with those based on continuous weak measurement schemes.
Notably, here the gravitational feedback enters entirely through the non-Hermitian jump operators, without modifying the unitary part of the dynamics. We show that this leads to a short-range gravitational back-reaction and permits decoherence rates below those of any model based on continuous weak measurement schemes.
We provide explicit examples, including the dynamics of a single particle and a rigid sphere, to illustrate the distinctive phenomenology of the model. Finally, we discuss the experimental testability of GPSL, highlighting both interferometric and non-interferometric strategies to constrain its parameters and distinguish it from competing models.
\end{abstract}

\maketitle

\section{Introduction}

Merging Quantum Mechanics and General Relativity is one of the most important challenges in physics. Most efforts have gone towards quantizing the gravitational field, producing remarkable frameworks such as String Theory and Loop Quantum Gravity; the complementary approach, which we denote as \enquote{Hybrid Classical-Quantum}, consists of keeping spacetime classical while matter is treated quantum mechanically. 
This latter approach has recently been gaining momentum~\cite{Tilloy2016CSLGravity,Tilloy2017LeastDecoherence,Tilloy2018GRWGravity,GaonaReyes2021GravitationalFeedback,Oppenheim2023PostQuantum,Oppenheim2023GravityTestsBounds} as a possible consistent theory while it has historically been seen as a stepping stone toward a fully quantum theory~\cite{Book_Hu2020SemiclassicalGravity}. 
In fact, the very nature of the gravitational field is still unclear: experiments have been proposed~\cite{Bose2017GravityExperiment,Marletto2017GravityExperiment,Lami2024GravityTesting,Angeli2025ProbingQuantumNatureGravity} to determine whether gravity is quantum, classical, or something else, but still have to be performed.

The first examples of a hybrid classical-quantum gravitational model are provided by {\it semiclassical gravity}, according to which classical gravity couples, as dictated by Einstein's equations, to the expectation value of the quantized stress-energy tensor~\cite{Book_Hu2020SemiclassicalGravity}.
This standard approach to semiclassical gravity is an active area of research~\cite{Husain2019FriedmannSchrodinger,Alonso2023HybridGeometrodynamics} as it allows to make, among others, cosmological predictions~\cite{Husain2021QuantumBackreaction}.
In the Newtonian limit, these semiclassical models are expected to reduce to the so-called Schr\"{o}dinger-Newton equation~\cite{Bahrami2014SchrodingerNewtonEquation}. However, if taken as a fundamental description of nature, they are problematic~\cite{Eppley1977NecessityQuantumGravity,Grobardt2022MakingSenseSemiclassicalGravity} as they suffer, among others, a severe conceptual issue: since the corresponding dynamics is nonlinear at the density matrix level~\cite{Bahrami2014SchrodingerNewtonEquation}, they allow for superluminal communication~\cite{Gisin1989StochasticDynamics}.

At the Newtonian level, this shortcoming is overcome by {\it stochastic hybrid models} based on the measurement plus feedback scheme~\cite{Book_Wiseman2009Measurement,Book_Jacobs2014MeasurementTheory}: quantum matter is suitably monitored over time, and the resulting stochastic measurement record is used to generate the classical gravitational field, which eventually interacts back with matter~\cite{Kafri2014LOCCGravity,Tilloy2016CSLGravity,Tilloy2018GRWGravity,Oppenheim2023PostQuantum,Tilloy2024HybridDynamics,Barchielli2024HybridDynamics}. By construction, there cannot be superluminal communication, because the resulting dynamics is linear at the density matrix level. 

These stochastic hybrid models differ among themselves essentially by what is measured and the way in which it is monitored~\cite{Kafri2014LOCCGravity,Tilloy2016CSLGravity,Tilloy2018GRWGravity,Tilloy2024HybridDynamics}.
Theoretical and experimental considerations suggest that, at the non-relativistic level, the monitored observable should be the (smeared) mass density~\cite{GaonaReyes2021GravitationalFeedback}.
Therefore, at the non-relativistic level and among models in which the mass density is measured, the main distinction is between those based on continuous weak measurements and those based on discrete Poissonian measurements. In the first class fall the Tilloy-Diósi (TD) model of Ref.~\cite{Tilloy2016CSLGravity} and all its variants; in the second class falls the model proposed by Tilloy in Ref.~\cite{Tilloy2018GRWGravity}.

The true challenge consists of how to formulate relativistic hybrid models, which ideally would represent an alternative to Quantum Gravity. 
New theoretical frameworks for such a formulation have been recently suggested~\cite{Oppenheim2022Constraints,Oppenheim2023PostQuantum,Oppenheim2023GravityTestsBounds}; however, they have been criticized because of the divergences contained in the measurement dynamics~\cite{Tilloy2024HybridDynamics,Diosi2024HybridRelativisticDynamicsDifficulties}.
As a matter of fact, it is well known that relativistic continuous weak measurements or, equivalently, relativistic models of spontaneous and continuous wave function collapse~\cite{Ghirardi1990RelativisticCSL,Pearle1990Toward,Adler2001RelativisticCSL,Bassi2003Dynamical,Bedingham2011Relativistic,Bassi2013Models}, the first building block of continuous hybrid models, present divergences that can be explained as a consequence of vacuum instabilities~\cite{Pearle1999Relativistic,Pearle2015RelativisticCSL,Myrvold2017RelativisticMarkovian,Jones2021MassCoupled,Diosi2024HybridRelativisticDynamicsDifficulties}\footnote{The relativistic model proposed in Ref.~\cite{Bedingham2011Relativistic} does not have this problem but the quantum system subject to the continuous weak monitoring is not a standard one.}.

Crudely speaking, the reason why vacuum instability leads to energy divergences is that, in Minkowski spacetime, the vacuum is invariant after the action of the entire Poincaré symmetry group. Therefore, any mixed state developing from it should have the same property~\cite{Myrvold2017RelativisticMarkovian}. However, the only other state with such a property is the completely mixed state, a state corresponding to an infinite temperature thermal state and thus an infinite-energy state. A generic curved spacetime does not possess the same symmetries of Minkowski spacetime. So, it is not obvious that a hybrid theory should also keep the vacuum stable. However, we deem it reasonable that it should, in some meaningful sense, at the level of quantum matter plus gravity (spacetime deformation). Accepting this, it also makes sense to assume that the non-relativistic limit of such a hybrid theory should maintain this property. This is why we argue that a Newtonian hybrid classical-quantum gravitational theory has to present a stable vacuum for being credible as the non-relativistic limit of the supposedly existing hybrid theory merging quantum mechanics and general relativity.

In the first approach to a relativistic spontaneous collapse model~\cite{Ghirardi1990RelativisticCSL}, the vacuum instability problem arises because, for the dynamics to be relativistically invariant, a Wiener noise in space and time is associated to the continuous weak measurement process, independently of the quantum state of the matter. This is violent enough to extract infinite energy~\cite{Ghirardi1990RelativisticCSL}; even when matter is absent, the measurement takes place, making vacuum unstable. A solution may be offered by discrete measurement processes, which by construction take place only when matter is present, and otherwise do not occur. 
In fact, there are already proposal of incomplete models of such kind: Ref.~\cite{Tumulka2006relativistic} develops a GRW-like model for relativistic but distinguishable and non-interacting particles, while Ref.~\cite{Tumulka2021relativistic} extends it to interacting particles (but still distinguishable and with a fixed number of them).

A non-relativistic discrete hybrid model was proposed in~\cite{Tilloy2018GRWGravity}, but works only for distinguishable particles. 
A generalization to identical particles and fields, a necessary step towards a relativistic formulation, was suggested in \cite{Piccione2023Collapse}, showing that the classical Newtonian field is recovered on average. A proper and thorough investigation of the properties of this new model, which we name Gravitational Poissonian Spontaneous Localization (GPSL), is the subject of this work.

We derive general properties of GPSL and compare its phenomenology with that of other hybrid models. As we will see, when applied to empty space, the dynamics give an \emph{exactly} vanishing gravitational field, since there are no collapses when there is no matter. Phenomenologically, the most striking difference between GPSL and continuous weak monitoring models is the absence of gravitational long-range decoherence in GPSL. Moreover, an explicit example shows how the GPSL model can present a spatial decoherence rate lower than the TD model, thus overcoming the so-called \enquote{Principle of Least Decoherence}~\cite{Tilloy2017LeastDecoherence}.
This implies that an experiment measuring a spatial decoherence rate lower than the bound set by the TD model would rule out all continuous weak monitoring models based on weak measurements but may not rule out GPSL.

Mathematically, we argue that all of the above properties mainly stem from the fact that, in contrast to the TD model and similar models, the gravitational interaction takes place entirely through the jump operators of the Lindbladian instead of being mainly due to the appearance of a new Hamiltonian term. The hybrid model structure in which gravitational feedback entails the appearance of a new Newton-like Hamiltonian term is often used to set generic bounds on hybrid models~\cite{Kafri2015BoundsNewtonianGravity,Angeli2025EntanglementHybridGravity,Angeli2025ProbingQuantumNatureGravity}. However, it is important to take into account that this structure may not be necessary, as GPSL seems to show.

The paper is structured as follows. In Sec.~\ref{Sec:GPSLModel}, we construct the GPSL model by defining first the measurement part and then the feedback part. In Sec.~\ref{Sec:GeneralProperties}, we explore some general properties of the GPSL model and, in Sec.~\ref{Sec:PerturbativeExpansion}, we show how to obtain a simplified and approximated version of it useful for practical calculations. Then, in Secs.~\ref{Sec:SingleParticle} and~\ref{Sec:RigidSphere}, we explicitly deal with two simple examples: an isolated particle and an isolated rigid sphere. Finally, in Sec.~\ref{Sec:Conclusions} we draw the conclusions of our paper.

\section{The GPSL Model\label{Sec:GPSLModel}}

The fundamental idea behind the GPSL model, which was suggested in Ref.~\cite{Piccione2023Collapse} but not investigated, is that the classical gravitational Newtonian field is sourced by the mass configuration generated by the results of spontaneous measurements of the (smeared) mass density operator; the measurements occur at random times and in random places, according to a specific probability distribution. Its empirical predictions are encoded in the following master equation:
\begin{multline}\label{eq:GPSL_MasterEquation}
\dv{t} \rho_t = -\frac{i}{\hbar}\comm{\hat{H}}{\rho_t} + 
\\
+\frac{\gamma}{m_0}\! \int\! \dd[3]{\bx}\! \prtq{U_G (\bx)\ghm(\bx)\rho_t \ghm(\bx) U_G^\dg (\bx) - \frac{1}{2}\acomm{\ghm^2(\bx)}{\rho_t}}\!,
\end{multline}
where $\hat{H}$ is the standard Hamiltonian of the system (without the gravitational part), $\ghm(\bx)$ is the square root of the smeared (over a radius $r_C$) mass density operator $\hmu_{r_C} (\bx)$, $\gamma$ is the collapse rate, $m_0$ is the proton mass, and $U_G (\bx)=\exp[-(i/\gamma\hbar) \hV (\bx)]$ is a unitary operator responsible for the gravitational dynamics, where $\hV (\bx) = - m_0 G\int \dd[3]{\by} \abs{\bx-\by}^{-1}\hmu_{r_C} (\by)$, with $G$ being the gravitational constant.
Notice how, in contrast to the models based on continuous weak measurements of the mass density (see Appendix~\ref{APPsec:TDmodelGeneral}), here the gravitational feedback does not entail the appearance of a new term in the Hamiltonian. The feedback is entirely contained in the modification of the jump operators\footnote{There may be a way to rewrite Eq.~\eqref{eq:GPSL_MasterEquation} in such a way that a new (and meaningful) Hamiltonian term appears, bringing out part of the gravitational feedback out of the jump operators. However, we did not find it and it may be that it is just not possible to do it.}.

The GPSL model presents two free parameters: the collapse rate $\gamma$, and the collapse radius $r_C$. The former quantifies how often a spontaneous measurement of $\hmu_{r_C} (\bx)$ takes place while the latter quantifies the smearing of the mass density operator, necessary to avoid divergences. The smearing is usually taken to be Gaussian, i.e.,
\begin{equations}
\hmu_{r_C} (\bx) 
&= (g_{r_C} * \hmu) (\bx)
= \int \dd[3]{\by} g_{r_C} (\bx-\by)\hmu(\by),
\\
g_{r_C} (\bx) &= \frac{\exp{-\bx^2/(2 r_C^2)}}{(2 \pi r_C^2)^{3/2}}.
\end{equations}
Other choices are possible but less common; therefore, we will henceforth assume a Gaussian smearing.

We now detail how Eq.~\eqref{eq:GPSL_MasterEquation} emerges, and the physics behind it. We separate the discussion in two parts: the first describes the (spontaneous) measurement process, and the second the gravitational feedback.

\subsection{The measurement part\label{Subsec:PSLModelNoGravity}}

The spontaneous measurement process which sources gravity in hybrid models is, in practice, a spontaneous collapse model. Spontaneous collapse models~\cite{Bassi2003Dynamical,Bassi2013Models,Bassi2023CollapseModels} modify standard quantum mechanics by adding collapse processes to it. These collapses are negligible for microscopic systems but dominant for macroscopic ones, thus explaining---at a fundamental level---the disappearance of quantum superpositions~\cite{Book_Bell2004Speakable}. The first model was proposed by Ghirardi-Rimini-Weber (GRW)~\cite{Ghirardi1986Unified}, assuming that every particle has a constant-in-time random probability of spontaneously localize in space. The GRW model, however, is not applicable to identical particles and fields without modifications; such generalizations exist~\cite{Tumulka2006spontaneous,Piccione2023Collapse} and we denote them as Poissonian Spontaneous Localization (PSL) models. The mass-proportional version of these models is the subject of this section. Other relevant spontaneous collapse models include the Continuous Spontaneous Localization (CSL) model~\cite{Ghirardi1990_CSL}, in its mass-dependent version~\cite{Pearle1994MassCSL}, and the Diósi-Penrose (DP) model~\cite{Diosi1987Universal,Diosi1989Models,Penrose1996gravity,Penrose2014Gravitization}.
Both of them are applicable to (non-relativistic) indistinguishable particles and fields, entail a continuous localization of the wavefunction, and can be used as the measurement part of the measurement plus feedback scheme proposed in Ref.~\cite{Tilloy2016CSLGravity}. Since spontaneous collapse models modify the \Schr dynamics, they can be experimentally falsified and bounds on their parameters have been investigated~\cite{Carlesso2019Collapse,Carlesso2022Present,Figurato2024DPEffectiveness,Piccione2025ExploringMassDependence}.

We briefly introduce the conceptual framework behind the PSL model~\cite{Piccione2023Collapse}. It is assumed that spacetime is filled with randomly distributed natural detectors (called collapse points) whose clicking probability is proportional to the average of the smeared mass density operator at their location and at the time of the spontaneous measurements. By considering the limit of high spacetime density of collapse points and their vanishing coupling with quantum matter, one obtains the PSL model, which is mathematically similar to the model of Ref.~\cite{Tumulka2006spontaneous}. The click of a collapse point corresponds to a spontaneous collapse of the wavefunction.

The resulting dynamics is given by the following stochastic \Schr equation~\cite{Piccione2023Collapse}:
\begin{multline}\label{eq:PSLDynamics}
\dd{\ket{\psi_t}}
= \prtqB{-\frac{i}{\hbar}\hat{H}\dd{t}+\int \dd[3]{\bx} \prt{\frac{\ghm(\bx)}{\sqrt{\ev{\hmu_{r_C} (\bx)}}}-1}\dd{N}_t (\bx)
	\\
	-\frac{\gamma}{2 m_0}\int \dd[3]{\bx} \prt{\hmu_{r_C} (\bx)-\ev{\hmu_{r_C} (\bx)}}\dd{t}
} \ket{\psi_t}.
\end{multline}
In the above equation, $\dd{N}_t (\bx)$ denotes the Poisson increment characterized by the following properties\footnote{In analogy to what done for the continuous models~\cite{Tilloy2016CSLGravity,GaonaReyes2021GravitationalFeedback}, one could want to consider a more general kind of Poissonian (but still Markovian) noise; one with spatial correlations between the Poisson infinitesimal increments at different spatial locations. However, this does not seem possible because, due to the noise being Poissonian, there cannot be two collapses at the same time.}: 
\begin{equations}\label{eq:PoissonProcessCharacterization}
\mbE\prtq{\dd{N}_t (\bx)} &= \frac{\gamma}{m_0}\ev{\hmu_{r_C} (\bx)}{\psi_t} \dd{t},\\ 
\dd{N}_t (\bx) \dd{t} &= 0,\\ 
\dd{N}_t (\bx)\dd{N}_t (\by) &= \delta (\bx-\by)\dd{N}_t (\bx).  
\end{equations}
A maybe useful analogy consists of considering Eq.~\eqref{eq:PSLDynamics} as a photo-detection stochastic Schr\"{o}dinger equation where a photon-counter is placed at every point in space and has a probability of clicking during an infinitesimal time $\dd{t}$ equal to $(\gamma/m_0)\ev{\hmu_{r_C} (\bx)}{\psi_t} \dd{t}$.
So, the probability of a collapse at point $\bx$ and at time $t$ is heuristically proportional to the amount of mass within a radius $r_C$ around $\bx$ at time $t$.
The master equation correspondent to the stochastic equation of Eq.~\eqref{eq:PSLDynamics} is 
\begin{equation}\label{eq:PSLMasterEquation}
\dv{t} \rho_t = -\frac{i}{\hbar}\comm{\hat{H}}{\rho_t} - \frac{\gamma}{2 m_0} \int \dd[3]{\bx} \comm{\ghm (\bx)}{\comm{\ghm (\bx)}{\rho_t}}.
\end{equation}

As anticipated, the collapse mechanism is practically negligible for microscopic systems while becomes important for larger ones. To see this in the PSL model, let us consider $N$ particles. We can write the smeared mass density operator as $\hmu_{r_C} (\bx) = \sum_{j=1}^N m_j g_{r_C} (\bx-\hbq_j)$, where $\hbq_j$ is the vector position operator for the $j$-th particle. In this case, the probability of a collapse happening \emph{anywhere} at any time between $t$ and $t+\dd{t}$ is given by
\begin{equation}
\mbE \prtq{\int \dd[3]{\bx} \dd{N_t} (\bx)} = \gamma \frac{M}{m_0} \dd{t},
\end{equation}
where $M = \sum_j m_j$ is the total mass of the $N$ particles. A typical range of values for the collapse rate is $\gamma \sim [10^{-9},10^{-3}]\rm{Hz}$~\cite{Piccione2025ExploringMassDependence}. Then, as long as the total mass of the system is microscopic, a collapse event is extremely rare within the usual time periods considered in experiments. The dynamics of the system is practically the same as that given by standard quantum mechanics, even for very delocalized states.

During a measurement process, in absence of collapses, the measured system would become entangled with a measurement apparatus. The possible measurement results are then encoded in macroscopically different states. Despite the fact that the dynamics of microscopic systems is basically unaltered by spontaneous collapses, this is not so for a macroscopic apparatus. In fact, the macroscopic apparatus is continuously kept in a definite macroscopic state (usually within a radius $r_A \ll r_C$). This is called \enquote{amplification mechanism}~\cite{Bassi2003Dynamical} and explains how measurements take place in a spontaneous collapse theory.

\subsection{The Feedback part\label{Subsec:FeedbackPart}}

The classical Newtonian gravitational field is added to the quantum dynamics following the strategy proposed in Ref.~\cite{Tilloy2018GRWGravity}: each collapse event  produces a gravitational impulse which, if needed, can be smeared both in space and time. 
(Here, we consider the case in which there is no smearing in time and the spatial smearing is transferred to the action of the gravitational field on quantum matter; in other words, the classical field couples to the operators $\hmu_{r_C} (\bx)$ instead of $\hmu (\bx)$.) 

Accordingly, the Newtonian gravitational potential becomes~\cite{Piccione2023Collapse}:
\begin{equations}\label{eq:PSLNewtonianField}
\Phi (\bx,t)\dd{t} &= \frac{m_0}{\gamma}\int \dd[3]{\by} \mcV (\bx,\by)\dd{N}_t (\by),
\\
\mcV (\bx,\by) &:= -\frac{G}{\abs{\bx-\by}},
\end{equations}
where $G$ is Newton's constant. 
Since in an infinitesimal time $\dd{t}$ there can only be at most one collapse, the above formula gives a zero potential $\Phi (\bx,t)\dd{t}=0$ if there is no collapse; if instead a collapse occurs  at $\by_0$ (i.e., $\dd{N}_t (\by)= \delta(\by-\by_0)$), it gives  the standard classical Newtonian potential associated to a (infinite) point mass located at that point.
Taking the average of Eq.~\eqref{eq:PSLNewtonianField} [see Eq.~\eqref{eq:PoissonProcessCharacterization}] over the Poisson process one gets
\begin{equations}
\mbE \prtq{\Phi (\bx,t)\dd{t}} 
&= \frac{m_0}{\gamma}\int \dd[3]{\by} \mcV (\bx,\by)\mbE \prtq{\dd{N}_t (\by)},
\\
&= \int \dd[3]{\by} \mcV (\bx,\by) {\ev{\hmu_{r_C} (\by)}}\dd{t},
\end{equations}
which is equal to the Newtonian potential generated by the mass distribution $\ev{\hmu (\by)}$, when considering distances much larger than $r_C$.

To obtain the dynamics of quantum matter interacting with the above classical gravitational potential in terms of a stochastic \Schr equation, one makes the substitution~\cite{Piccione2023Collapse} $\ghm (\bx) \rightarrow U_G (\bx) \ghm (\bx)$ in Eq.~\eqref{eq:PSLDynamics}, where:
\begin{equations}
U_G (\bx) 
&= \exp{-\frac{i}{\gamma\hbar} \hV (\bx)},
\\
\hV (\bx) 
&= m_0 \int \dd[3]{\by} \mcV (\bx,\by) \hmu_{r_C} (\by),
\end{equations}
$U_G (\bx)$ being the unitary gravitational impulse associated to a collapse located in $\bx$. Explicitly, one gets:
\begin{multline}\label{eq:PSLDynamicsWithGravity}
\dd{\ket{\psi_t}}
= \prtqB{-\frac{i}{\hbar}\hat{H}\dd{t} + \int \dd[3]{\bx} \prt{\frac{U_G (\bx
			) \ghm (\bx)}{\sqrt{\ev{\hmu_{r_C} (\bx)}}}-1}\dd{N}_t (\bx)
	\\
	-\frac{\gamma}{2 m_0}\int \dd[3]{\bx} \prt{\hmu_{r_C} (\bx)-\ev{\hmu_{r_C} (\bx)}}\dd{t}
}\ket{\psi_t}.
\end{multline}
Let us notice that $\hV (\bx)$ is a \enquote{sum} of smeared mass density operators, so $\comm{U_G (\bx)}{\hmu_{r_C} (\by)} = \comm{U_G (\bx)}{\ghm(\by)} = 0$. Physically, this implies that it does not matter whether the gravitational feedback takes place before or after the collapse.
The Lindblad equation associated to Eq.~\eqref{eq:PSLDynamicsWithGravity} is Eq.~\eqref{eq:GPSL_MasterEquation}.

For $N$ particles, the stochastic equation simplifies because $\hmu_{r_C} (\bx) = \sum_{j=1}^N m_j g_{r_C} (\bx-\hbq_j)$ so that $\int\dd[3]{\bx} \ghm^2 (\bx) = \sum_j m_j = M$, the total mass of the system. Therefore, one has
\begin{equations}\label{eq:PSLDynamicsWithGravityParticles}
&\dd{\ket{\psi_t}}
= \prtq{\int \dd[3]{\bx} \prt{\frac{U_G (\bx
			) \ghm (\bx)}{\sqrt{\ev{\hmu_{r_C} (\bx)}}}-1}\dd{N}_t (\bx)} \ket{\psi_t},\\
&\dv{t} \rho_t = -\frac{\gamma}{m_0} \prtqB{M\rho_t -\int \dd[3]{\bx}
	U_G (\bx)\ghm (\bx)\rho_t\ghm (\bx)U_G^\dg (\bx)},\\
&U_G (\bx) 
= e^{i \sum_k r_k f(\hbq_k-\bx)},
\end{equations}
where we defined $r_k \equiv (G m_0 m_k)(\gamma \hbar)^{-1}$, $f(\bx)=\abs{\bx}^{-1}\erf(\abs{\bx}/(r_C\sqrt{2}))$, and we omitted the standard Hamiltonian term for conciseness.

In the rest of the paper, we will often make comparisons with the gravitational models based on continuous weak monitoring of the (smeared) mass density operator, summarized in Appendix~\ref{APPsec:TDmodelGeneral}. In these models, the quantity sourcing the Newtonian field is~\cite{Tilloy2016CSLGravity} $\mu_t (\bx) = \ev{\hmu_{r_C} (\bx)} + \delta \mu_t (\bx)$, where $\delta \mu_t (\bx)$ represents a white-noise in time with a potentially non-trivial spatial correlator. This means that there are, continuously in time, measurement results for every point in space (also when $\ev{\hmu_{r_C} (\bx)}=0$), which can give back a negative mass reading. This feature suggests that the vacuum may not be stable in a relativistic context; this, as discussed in the introduction, is a key problem of relativistic models containing continuous weak measurements~\cite{Myrvold2017RelativisticMarkovian}. In the GPSL model, instead, every collapse corresponds to the \enquote{detection} of a positive mass and there are no collapses when there is no mass. Therefore, these potentially problematic features are avoided by construction.

\section{General Properties of the Model\label{Sec:GeneralProperties}}

We now explore the most relevant properties of the GPSL dynamics encoded in Eq.~\eqref{eq:GPSL_MasterEquation}, and compare them to those of the continuous weak monitoring models.

\subsection{Field Fluctuations}
In Appendix~\ref{APPSec:FieldFluctuations}, we compute the spatial covariance $\mbE [\Phi (\bx,t) \dd{t}\Phi (\by,t) \dd{t}]$ of the Newtonian field in the GPSL and continuous weak monitoring models. The variance of the fields at a point $\bx$ is obtained by setting $\bx=\by$. For the GPSL model, we get
\begin{multline}
\mbE \prtq{\Phi (\bx,t)\dd{t}\Phi (\by,t)\dd{t}}
=\\=
\frac{m_0}{\gamma} \int \dd[3]{\bz}\mcV(\bx,\bz)\mcV(\by,\bz) \ev{\hmu_{r_C} (\bz)}_t\dd{t},
\end{multline}
while for a generic continuous weak monitoring model we have
\begin{equation}
\mbE \prtq{\Phi (\bx,t)\dd{t}\Phi (\by,t)\dd{t}}
=
\frac{1}{4}(\mcV \circ \gamma_C^{-1} \circ \mcV)(\bx,\by)\dd{t},
\end{equation}
where
\begin{equation}
(f \circ g)(\bx,\by) =
\int \dd[3]{\bz} f(\bx,\bz)g(\bz,\by),
\end{equation}
and $\gamma_C^{-1} (\bx,\by)$ is the inverse of the noise correlation kernel. 
Notice how the covariance, in the GPSL case, is vanishing when $\ev{\hmu_{r_C} (\bz)}_t=0$. Indeed, this happens because there are no collapses when there is no mass, and the resulting field is always zero with zero fluctuations. On the contrary, in the continuous weak monitoring models, the covariance is independent of the mass content in space. Moreover, without a smearing of $\mcV (\bx,\by)$, the variance of the continuous weak monitoring models based on the CSL or DP spontaneous collapse models is divergent (see Appendix~\ref{APPSec:FieldFluctuations}) while, in the GPSL case, this smearing is not needed.

\subsection{Short Range Gravitational Back-Reaction\label{Subsec:ShortRangeDecoherence}}

We want to show that the gravitational feedback of the GPSL model is short-ranged in the sense that it is completely negligible when considering spatial superpositions at distances much larger than $r_C$ or than the system's dimensions (whichever is higher). This is easily understood when considering $N$ particles so that [See Eq.~\eqref{eq:PSLDynamicsWithGravityParticles}]
\begin{multline}
\dv{t} \rho_t (x,y) = -\frac{\gamma}{m_0}M \rho_t (x,y)\prtqB{1 
	+\\- 
	\frac{1}{M}\int \dd[3]{\bz} e^{i\sum_k r_k [f(\bx_k-\bz)-f(\by_k-\bz)]} 
	\times \\ \times 
	\sqrt{\sum_{j,j'} m_{j}m_{j'} g_{r_C} (\bx_j - \bz) g_{r_C} (\by_{j'} - \bz)}},
\end{multline}
where $x$ and $y$ denote the collective coordinates of the $N$ particles. When the entire system is in a superposition of two locations so that $g_{r_C} (\bx_j - \bz) g_{r_C} (\by_{j'} - \bz) \simeq 0$, $\forall \bx_j,\by_{j'}$, the effect of the gravitational back-reaction is negligible.

\begin{figure}[t]
\centering
\includegraphics[width=0.45\textwidth]{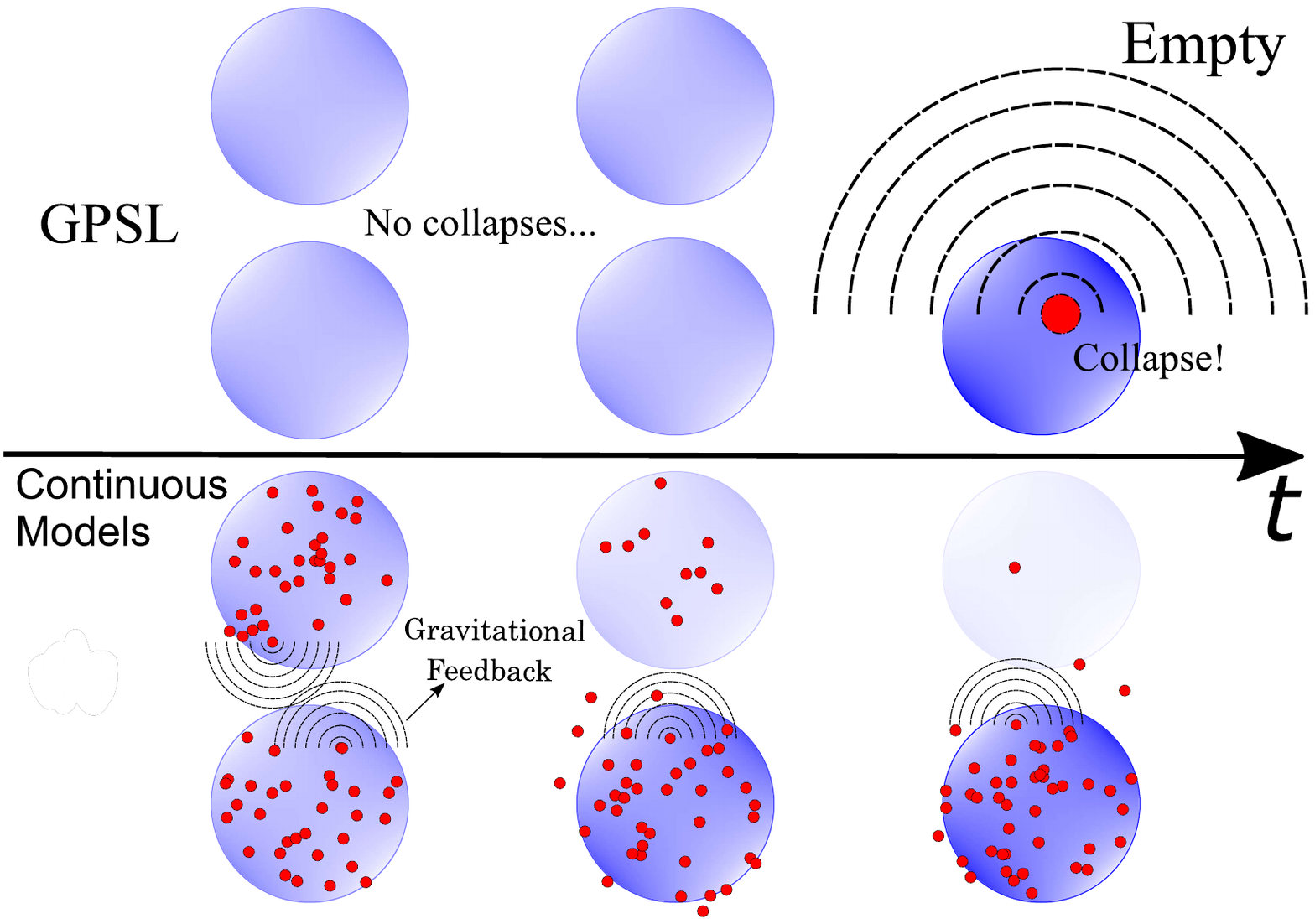}
\caption{Pictorial explanation of the localization dynamics with gravitational back-reaction, starting from a balanced superposition of a particle in two different locations. 
	In the GPSL model (above the timeline), the superposition remains balanced and the gravitational field is always zero until the instantaneous collapse (represented by the red circle) occurs. When it does, the body immediately localizes and the instantaneously generated field can only have effect where the body is located, thus explaining the absence of long-range decoherence due to the gravitational back-reaction. In the continuous weak monitoring models (below the timeline), the spatial superposition is gradually suppressed by the continuous weak measurements (represented by the small red circles) and the gravitational back-reaction acts during this process, giving rise to additional long-range decoherence due to the gravitational back-reaction.
}
\label{fig:ShortRangeExplanation}
\end{figure}

The effect predicted above can be explained as follows (see Fig.~\ref{fig:ShortRangeExplanation}). When a system is in a superposition of distant locations, any collapse event will immediately locate it to one of the possible locations. The gravitational field that is generated as consequence of this collapse is non-zero but all locations different from where the collapse occurred are now empty, so there is no effect on the matter except for where matter is actually localized. The situation is significantly different in the continuous weak monitoring models, where the gravitational back-reaction leads to long-range decoherence as opposed to the short-range one of the GPSL model. In the continuous weak monitoring case, this happens because the gravitational feedback is in place while the wavefunction is continuously collapsing so that the field now interacts with multiple non-empty locations. We will see this general prediction confirmed in the particular cases of a single particle (Sec.~\ref{Sec:SingleParticle}) and of a rigid sphere (Sec.~\ref{Sec:RigidSphere}). 

\subsection{Average Momentum Conservation and Average Forces\label{Subsec:TotalMomentumConservation}}

The GPSL model does not select any special location in space. Therefore, translation invariance of the model is expected for isolated systems. Indeed,
when considering an ensemble of $N$ particles, Eq.~\eqref{eq:GPSL_MasterEquation} is evidently translation invariant as all the operators are functions of position operators that are integrated over space. However, for open quantum systems, a symmetry does not imply a conservation law~\cite{Manzano2014SymmetryOpenSystems,Albert2014SymmetriesLindblad,Albert2019AsymptoticsOfQuantumChannels,Cirstoiu2020RobustnessNoether}. In fact, already the collapse dynamics alone should induce diffusion~\cite{Donadi2023DiffusiveCollapse} so that the total momentum operator of an isolated system is not conserved. 
Here, we show that the \emph{average} of the total momentum of an isolated system is conserved in the GPSL model.

The dynamics of an isolated system composed of $N$ particles can be described as follows:
\begin{equation}
\dv{t} \rho_t
= -\frac{i}{\hbar}\comm{\frac{\hbP^2}{2 M} + \hat{H}_r}{\rho_t}
+ \mcL_{\rm GPSL} (\rho_t),
\end{equation}
where $\hbP$ is the total momentum operator, $M$ the total mass of the system, $\hat{H}_r$ the internal Hamiltonian (which, by assumption, commutes with all operators referring to the center of mass), and $\mcL_{\rm GPSL}$ is the Lindblad super-operator associated to the GPSL model. Changes of the total momentum can only come from the Lindbladian, but showing that $\tr\prtg{\hbP \mcL_{\rm GPSL} (\rho_t)}=0$ is not as easy as it may seem. In the PSL model (without gravity), this can be easily verified by exploiting the unitary unraveling of Eq.~\eqref{eq:PSLMasterEquation}~\cite{Piccione2025ExploringMassDependence}. Similarly, a unitary unraveling can be used to easily show the conservation of the average total momentum in any continuous weak monitoring model (see Appendix~\ref{APPSubsec:UnitaryUnravelingTDModels}).

To show the total momentum conservation in the GPSL model, we found it convenient to smear the feedback response in time by an amount $\tau$ so that we can separate the gravitational feedback from the spontaneous collapse dynamics. The details are given in Appendix~\ref{APPsec:ConservationAverageMomentum}. The final result is that the average force (over state and noise) exerted by the $j$-th particle on the $k$-th particle is given by
\begin{multline}\label{eq:PairwiseNewtonianForceGPSL}
\bar{\mathbf{F}}^{k,j} 
=
G m_k m_j
\int \dd[3]{\bx}\dd[3]{\bz_j}\dd[3]{\bz_k} \rho^{(j,k)}_{t} (\bz_j,\bz_k;\bz_j,\bz_k) 
\times \\ 
g_{r_C} (\bx+\bz_k-\bz_j) \frac{\bx}{\abs{\bx}^2}
\prtq{\frac{1}{\abs{\bx}}\erf \prt{\frac{\abs{\bx}}{r_C\sqrt{2}}}-4 \pi r_C^2 g_{r_C} (\bx)},
\end{multline}
where $\rho^{(j,k)}_{t} (\bz_j,\bz_k;\bz_j',\bz_k')$ is the reduced density matrix of the $j$-th and $k$-th particles in position representation. Notably, the above formula reduces to Newton's formula for the gravitational force when $r_C \rightarrow 0$ and $\rho^{(j,k)}_{t} (\bz_j,\bz_k;\bz_j,\bz_k) =\delta (\bz_j-\bx_{j})\delta (\bz_k-\bx_{k})$. Moreover, using the spherical symmetry of $g_{r_C} (\bx)$, a simple change of variable $\bx \rightarrow -\bx$ shows that $\bar{\mathbf{F}}^{k,j} = -\bar{\mathbf{F}}^{j,k}$ so that, at the average level, pairs of forces neutralize and it follows that the average total momentum is conserved. We remark, however, that the average energy is not constant. It is actually expected to increase, as this is required for any collapse model that is effective in localizing superpositions of matter~\cite{Donadi2023DiffusiveCollapse}.

\section{Perturbative Expansion of the GPSL Master Equation\label{Sec:PerturbativeExpansion}}

Eq.~\eqref{eq:GPSL_MasterEquation} is difficult to manage for actual computations. The master equation becomes more tractable if the unitary operator $U_G (\bx)$ is expanded perturbatively. We first motivate why this is possible and then we perform the expansion.

Let us consider an ensemble of particles so that $\hmu_{r_C} (\bx) = \sum_k m_k g_{r_C} (\bx- \hbq_k)$ where $m_k$ is the mass of the $k$-particle and $\hbq_k$ its position operator. Then, we have that
\begin{equations}\label{eq:UnitaryFeedbackSingleParticlesDecomposition}
\hV (\bx) &= -G m_0 \sum_k m_k \frac{1}{\abs{\hbq_k-\bx}}\erf\prt{\frac{\abs{\hbq_k-\bx}}{r_C\sqrt{2}}},\\
U_G (\bx) 
&= e^{i \sum_k r_k f(\hbq_k-\bx)},
\end{equations}
where we recall that $r_k \equiv (G m_0 m_k)(\gamma \hbar)^{-1}$ and $f(\bx)=\abs{\bx}^{-1}\erf(\abs{\bx}/(r_C\sqrt{2}))$. To check if the perturbative expansion of $U_G (\bx)$ for a given point $\bx$ can be done, we can estimate $\sum_k r_k f(\hbq_k-\bx)$ where $k$ goes through all the particles contained in a ball of radius $r_C$ centered at $\bx$. Denoting by $N_T$ the number of particles of type $T$ contained in the ball, we get that
\begin{equation}
\sum_k r_k f(\hbq_k-\bx)
\leq \sum_T N_T r_T \max_{\bx} f(\bx)
= \sum_T \frac{N_T r_T\sqrt{2/\pi}}{r_C}.
\end{equation}
So, the expansion can be made if $\sum_T N_T r_T \ll r_C$. To see that this is usually the case let us consider a system with a standard particle density of approximately $n_0 \sim 10^{23}/\rm{cm}^3$. The smallest considered value for the collapse rate is $\gamma \sim 10^{-9} \textrm{Hz}$~\cite{Piccione2025ExploringMassDependence}, while a typical value for $r_C$ is $r_C \sim 10^{-7} \textrm{m}$. Then, there are around $N_T \sim n_0 r_C^3 \sim 10^8$ particles\footnote{We consider all particles of the same kind for simplicity.} in a ball of radius $r_C \sim 10^{-7} \textrm{m}$. Taking $m_T \sim m_0$ we then get $N r_T/r_C \sim 10^{-6}$, which justifies the approximation even for very heavy molecules. So, the above approximation holds unless the particle density is extremely high. For a Neutron star, for example, one would have $N_T \sim 10^{23}$ and $N_T r_T/r_C \sim 10^{9}$, so the above approximation would not be valid.
Finally, keeping only the terms up to second order one gets
\begin{equation}\label{eq:UnitaryGravitationalImpulsePerturbativeExpansion}
U_G (\bx) \simeq 1 - \frac{i}{\hbar\gamma}\hV (\bx) - \frac{1}{2\hbar^2\gamma^2} \hV^2 (\bx).
\end{equation}

Let us now use Eq.~\eqref{eq:UnitaryGravitationalImpulsePerturbativeExpansion} to write a simpler master equation. First, we have that
\begin{multline}\label{eq:ExpansionOfUnitaryGravitationalFeedbackSecondOrder}
U_G (\bx) \rho_t U_G^\dg (\bx) 
\simeq \\ \simeq
\prtq{1 - \frac{i\hV (\bx)}{\hbar\gamma} - \frac{\hV^2 (\bx)}{2\hbar^2\gamma^2} }
\rho_t
\prtq{1 + \frac{i\hV (\bx)}{\hbar\gamma} - \frac{\hV^2 (\bx)}{2\hbar^2\gamma^2} }
\simeq \\ \simeq
\rho_t - \frac{i}{\hbar \gamma }\comm{\hV (\bx)}{\rho_t} - \frac{1}{2 \hbar^2 \gamma^2} \comm{\hV(\bx)}{\comm{\hV(\bx)}{\rho_t}}.
\end{multline}
Then, substituting this in Eq.~\eqref{eq:GPSL_MasterEquation} we get
\begin{multline}\label{eq:PerturbativeGPSL_MasterEquation}
\dv{t} \rho_t = -\frac{i}{\hbar}\frac{1}{m_0}\int \dd[3]{\bx}\comm{ \hV (\bx)}{\ghm (\bx)\rho_t\ghm (\bx)}
+\\+
\frac{\gamma}{m_0} \int \dd[3]{\bx} \mcD_{\ghm (\bx)} (\rho_t)
+\\+
\frac{1}{m_0 \hbar^2 \gamma} \int \dd[3]{\bx} \mcD_{\hV (\bx)} \prt{\ghm (\bx)\rho_t\ghm (\bx)},
\end{multline}
where we introduced the notation $\mcD_{\hA} (\rho) = \hA \rho \hA^\dg -(1/2)\{ \hA^\dg \hA,\rho \}$ to save space. Eq.~\eqref{eq:PerturbativeGPSL_MasterEquation} is not in Lindblad form because of the approximations made to obtain it. However, it is still trace-preserving and its complete positivity (at second order in $(\hbar \gamma)^{-1}\hV (\bx)$) should hold when the conditions for using Eq.~\eqref{eq:UnitaryGravitationalImpulsePerturbativeExpansion} are satisfied.

In Eq.~\eqref{eq:PerturbativeGPSL_MasterEquation}, we see that two new terms appear with respect to Eq.~\eqref{eq:PSLMasterEquation}. One is a unitary-like term and one is a dissipator-like term. The latter is akin to the additional decoherence term that appears in the continuous weak monitoring models~\cite{Tilloy2016CSLGravity}. In particular, as in the continuous weak monitoring model based on the CSL spontaneous collapse model, the new dissipative term contains a factor $\gamma^{-1}$, potentially allowing for experimental lower bounds of $\gamma$ once $r_C$ is fixed.
The former, instead, should be responsible for inducing the expected Newtonian behavior of single particles in a gravitational field generated by the other masses. To show this, let us consider a particle of mass $m$, at distances much larger than $r_C$ from all other masses. Moreover, let us consider a state $\rho_t$ obeying the equation
\begin{equation}
\dv{t} \rho_t = -\frac{i}{\hbar}\frac{1}{m_0}\int \dd[3]{\bx}\comm{ \hV (\bx)}{\ghm (\bx)\rho_t\ghm (\bx)},
\end{equation}
and such that $\rho_t = \rho_t^{(p)} \otimes \rho_t^{(\mu)}$, where $\rho_t^{(p)}$ is the reduced state of the particle of interest and $\rho_t^{(\mu)}$ is the state of all the other masses.
In Appendix~\ref{APPsec:ReducedParticleDynamics}, we show that
\begin{equation}\label{eq:SingleParticleGravitationalUnitaryReducedDynamics}
\dv{t} \rho_t^{(p)} \simeq -\frac{i}{\hbar}\comm{\int \dd[3]{\bx}\frac{-G m \Tr \{ \hmu (\bx) \rho_t^{(\mu)}\}}{\abs{\hbq_p-\bx}}}{\rho_t^{(p)}},
\end{equation}
where $\Tr \{ \hmu (\bx) \rho_t^{(\mu)}\}$ is the average density mass distribution of all the other masses.
This is indeed the expected behavior as the Hamiltonian term appearing inside the commutator corresponds to the Newtonian potential energy between the particle of mass $m$ and the mass distribution $\Tr \{ \hmu (\bx) \rho_t^{(\mu)}\}$.

\section{Single Particle Dynamics\label{Sec:SingleParticle}}

We now explore the dynamics of an isolated particle and compare it to two continuous weak monitoring models: one based on the CSL spontaneous collapse model and one based on the DP one. In doing so, we will neglect the standard quantum dynamics of the particle as we are interested in the spatial decoherence due to the gravitational back-reaction.

When neglecting the standard Hamiltonian, the density matrix of a single particle of mass $m$ in position representation evolves as follows:
\begin{equations}\label{eq:SingleParticleDecoherence}
\dv{t} \rho_t (\bx,\by) &= -\Gamma (\bx,\by)\rho_t(\bx,\by),\\
\Gamma (\bx,\by) &= \frac{\gamma m}{m_0}\prtqB{1+\\
	-\int \dd[3]{\bz}\ &e^{i r_p [f(\bx-\bz)-f(\by-\bz)]}\sqrt{g_{r_C}(\bx-\bz)g_{r_C}(\by-\bz)}},
\end{equations}
where we recall that $r_p = (G m_0 m)(\gamma \hbar)^{-1}$ and $f(\bx)=\abs{\bx}^{-1}\erf(\abs{\bx}/(r_C\sqrt{2}))$. In Appendix E.2 of Ref.~\cite{Piccione2023Collapse}, it is shown that $\Gamma (\bx,\by)$ is real so that the only effect of the collapse dynamics plus gravitational back-reaction is spatial decoherence.

\begin{figure}[t]
\centering
\includegraphics[width=0.45\textwidth]{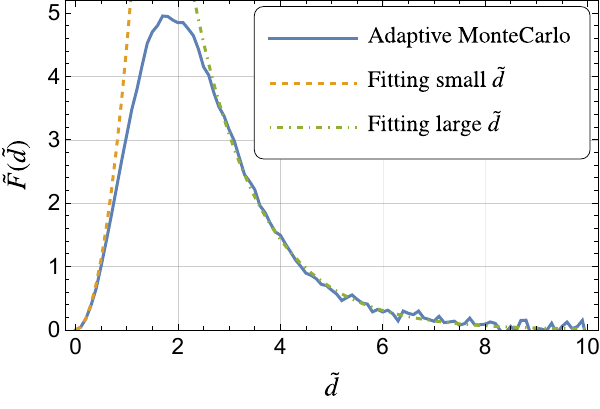}
\caption{The plot shows a numerical evaluation of $\tl{F} (\tl{d})$ using the \enquote{Adaptive MonteCarlo} built-in method in Mathematica. We then fitted the resulting curve with $4.49 \times \td^2$ on the left and $2.1\times \exp{-(\td - 3.5)/1.3}$ on the right.}
\label{fig:plotFtilde}
\end{figure}

For the same reason as in Sec.~\ref{Sec:PerturbativeExpansion}, we can expand the exponential inside the integral to second order so to isolate the gravitational contribution\footnote{The resulting expression could have been obtained by calculating $\Gamma (\bx,\by)$ from Eq.~\eqref{eq:PerturbativeGPSL_MasterEquation} and noticing that the unitary-like term vanishes.}. The resulting expression contains an integral computed in Appendix~\ref{APPsec:SingleParticleDecoherencePSL}. The final result is
\begin{equation}\label{eq:SingleParticlePSLGravitationalDecoherence_Evaluated}
\Gamma (\bx,\by) = 
\gamma \frac{m}{m_0}\prtq{1-e^{-\td^2/2}
	+\frac{r_p^2}{r_C^2}\frac{e^{-\td^2/2}}{2\pi^4} \tl{F}(\td)},
\end{equation}
where $\td \equiv \abs{\bx-\by}/(2 r_C)$ and $\tl{F}(\td)$ is a function that has to be evaluated numerically; $\tl{F}(\td)$ is plotted and fitted in Fig.~\ref{fig:plotFtilde}. In the above equation, the first two terms between square parentheses are due to the collapse dynamics alone and would be present even in absence of the gravitational feedback. The last term is instead due to the gravitational feedback and goes to zero more than exponentially fast because $\tl{F}(\td)$ goes to zero exponentially fast for $\td \gg 1$. For $\td \ll 1$, we have that $\tl{F}(\td) \simeq 4.49\times \td^2$.

We can compare the GPSL decoherence rate with those of the continuous weak monitoring models based on the CSL and DP spontaneous collapse models (see Appendix~\ref{APPSubsec:TDModelSingleParticlesDecoherenceRate}):
\begin{equations}\label{eq:TDModelsDecoherenceRatesSingleParticle}
\Gamma^{\rm CSL}(\bx,\by)
&= \frac{\gamma_{\rm CSL}}{(4\pi r_C^2)^{3/2}}\prt{\frac{m}{m_0}}^2
\prtq{1-e^{-\td^2}}+\\
+\frac{\pi r_C}{\gamma_{\rm CSL}} &\frac{m_0^2 m^2 G^2}{\hbar^2} \prtq{\prt{\td +\frac{1}{2\td}}\erf(\td) -\frac{2-e^{-\td^2}}{\sqrt{\pi}}},
\\
\Gamma^{\rm DP}(\bx,\by)
&= \frac{G m^2}{\hbar r_C \sqrt{\pi}}\prtq{1 - \frac{\sqrt{\pi}}{2\td}\erf\prt{\td}}.
\end{equations}
In the CSL case, the first term comes from the spontaneous collapse dynamics while the second one from the gravitational back-reaction\footnote{Notice, though, that with the typical values of $r_C \sim 10^{-7} \rm{m}$ and $\gamma_{\rm CSL} \sim 10^{-36} \rm{m}^{3} \rm{s}^{-1}$, the ratio between the pre-factors of the collapse and gravitational decoherence gives $\pi(4\pi)^{3/2} G^2 m_0^4 r_C^4 (\hbar \gamma_{\rm CSL})^{-2} \sim 4.4 \times 10^{-14}$ so that the gravitational contribution would be comparable with the collapse one when the superposition is at distances of the order of $10^{7} \rm{m}$.}. In the DP case, the spontaneous collapse and the gravitational back-reaction both account for half of the decoherence rate~\cite{Tilloy2017LeastDecoherence}. The long-range behavior of the gravitational back-reaction is strikingly different in the three cases. In the GPSL case, the gravitational back reaction is rapidly suppressed for $\td \gg 1$, in the DP case it saturates, and in the CSL case it grows linearly with $\td$. The fact that the decoherence rate is short-ranged in the GPSL model while it is not in the continuous weak monitoring models is a consequence of the general argument presented in Sec.~\ref{Subsec:ShortRangeDecoherence}.

Before continuing, let us briefly make a comment on the so-called \enquote{Principle of Least Decoherence}. In Ref.~\cite{Tilloy2017LeastDecoherence}, Tilloy and Diósi proposed a way to choose a continuous spontaneous collapse models among all the possible ones. They argued that one should choose the collapse model that, once the gravitational back-reaction is added, leads to the lowest possible spatial decoherence rate; this requirement singles out the Diósi-Penrose model. However, the calculations of Ref.~\cite{Tilloy2017LeastDecoherence} only apply to collapse dynamics based on continuous weak measurements. We now explicitly show that, in theory, the GPSL model can lead to a lower decoherence rate than the TD~\cite{Tilloy2017LeastDecoherence}. This is easily done by considering a large spatial superposition such that $\abs{\bx-\by} \rightarrow +\infty$. Then, one immediately gets\footnote{Notice that we are not assuming $r_p \ll r_C$ for getting this result.} [cf. Eqs.~\eqref{eq:SingleParticleDecoherence} and~\eqref{eq:TDModelsDecoherenceRatesSingleParticle}]
\begin{equation}
\Gamma^{\rm GPSL} \rightarrow \gamma\frac{m}{m_0},
\quad
\Gamma^{\rm DP} \rightarrow \frac{G m^2}{\hbar r_C \sqrt{\pi}}.
\end{equation}
Mathematically speaking, one can obtain that $\Gamma^{\rm GPSL}<\Gamma^{\rm DP}$ by choosing $\gamma$ small enough or $m$ large enough. In particular, for any value of $\gamma$ and $r_C$, there exists a large enough mass $m$ such that $\Gamma^{\rm GPSL}<\Gamma^{\rm DP}$.
Conversely, for any value of $\gamma$ there is also a small enough $m$ that $\Gamma^{\rm GPSL}>\Gamma^{\rm DP}$. For example, using values such as $\gamma \sim 10^{-9} \textrm{Hz}$ and $r_C \sim 10^{-7} \textrm{m}$, one has that $\Gamma^{\rm GPSL}<\Gamma^{\rm DP}$ for both electrons, protons, and even much heavier particles. Indeed, we are not claiming that GPSL is, generally speaking, harder to experimentally falsify than TD, but only that there are situations in which one can have $\Gamma^{\rm GPSL}<\Gamma^{\rm DP}$, while, even mathematically, it can never happen that $\Gamma^{\rm CSL}<\Gamma^{\rm DP}$. As a final comment, one may wonder if a general Principle of Least Decoherence for GPSL can be satisfied. The answer is negative because the value of $\gamma$ minimizing Eq.~\eqref{eq:SingleParticlePSLGravitationalDecoherence_Evaluated} depends on the particle's mass and the superposition distance.

\section{Macroscopic Rigid Body Dynamics\label{Sec:RigidSphere}}

In this section, as previously done for the single particle, we investigate the decoherence rate of a rigid body, in particular of a rigid sphere of radius $R \gg r_C$. We choose a sphere as it is the simplest rigid body on which (approximate) analytical results can be obtained.

In Ref.~\cite{Piccione2025ExploringMassDependence}, we have shown how the action of the operator $\hmu_{r_C} (\bx)$ decouples at the level of center of mass and internal coordinates under the assumption of a rigid body whose orientation in space is fixed. More than this, if the rigid body is of constant density, is characterized by a typical length much higher than $r_C$, $r_C$ is big enough to contain many atoms/molecules when considering balls of radius $r_C$ within the body, and the rigid body's surface varies on scales much larger than $r_C$, then the smeared mass density operator takes the following form\footnote{It may be useful to recall that $F(z)=[1+\erf (z/\sqrt{2})]/2$ is the cumulative distribution function of the standard normal distribution.}:
\begin{equation}\label{eq:RigidBodyMassDensityOperator}
\hmu_{r_C} (\bx) = \varrho_{r_C}(\bx-\hbQ) = \frac{M}{V} \prtq{\frac{1}{2} + \frac{1}{2}\erf \prt{\frac{d(\bx-\hbQ)}{r_C\sqrt{2}}}},
\end{equation}
where $M$ is the total mass of the system, $\hbQ$ is the center-of-mass position operator, while $d(\bx)$ is the (minimal in absolute value) distance between $\bx$ and the rigid body's surface with $d(\bx)>0$ if $\bx$ is within the body and $d(\bx)<0$ if it is outside of it.

To analyze the rigid body dynamics we will use Eq.~\eqref{eq:PerturbativeGPSL_MasterEquation}.
In the position representation, the unitary-like term gives the following contribution:
\begin{equation}
\dv{t} \rho_t^{\rm CM} (\bX,\bY) = -\frac{i}{\hbar}\Gamma_U^{\rm CM} (\bX,\bY) \rho_t^{\rm CM} (\bX,\bY),
\end{equation}
with
\begin{multline}
\Gamma_U^{\rm CM} (\bX,\bY) 
=\\
+\int \dd[3]{\bz}\dd[3]{\bz'} \mcV(\bz,\bz')\varrho_{r_C}(\bz')\sqrt{\varrho_{r_C} (\bz)\varrho_{r_C} (\bz+\bD)}
\\
-\int \dd[3]{\bz}\dd[3]{\bz'} \mcV(\bz,\bz')\varrho_{r_C}(\bz')\sqrt{\varrho_{r_C} (\bz)\varrho_{r_C} (\bz-\bD)}
\end{multline}
where $\bD = \bX-\bY$, and $\bX(\bY)$ denote center of mass position. When the two integrals give the same result, for $\bD$ and $-\bD$, one has $\Gamma_U^{\rm CM} (\bX,\bY) = 0$. This is clearly the case for many shapes with a certain amount of symmetries (such as a sphere).

\begin{figure}[t]
\centering
\includegraphics[width=0.45\textwidth]{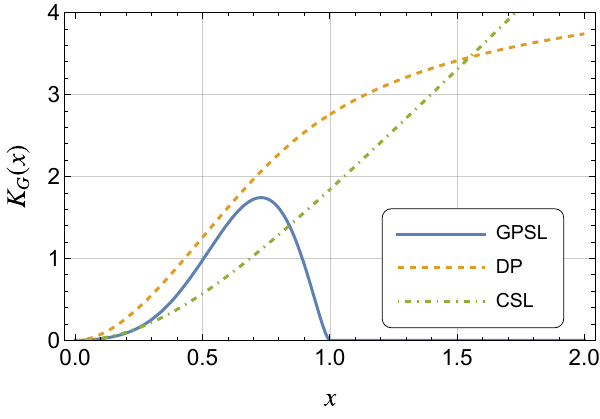}
\caption{The plot shows the functions $K_{G}^{\rm GPSL} \prt{x}$, $K_{G}^{\rm DP} \prt{x}$, and $K_{G}^{\rm CSL} \prt{x}$.}
\label{fig:plotDecoherenceRatesRigidSpheres}
\end{figure}

Next, we study the decoherence-like term. To estimate the decoherence rate we make use of the approximation\footnote{Notice that if we were interested in the energy increase of the rigid body, we could not use this approximation but should instead use a more sophisticated one, as done in Ref.~\cite{Piccione2025ExploringMassDependence}.} 
$\varrho(\bx) = (M/V) \chi_V (\bx)$, where $M$ is the mass of the rigid body, $V$ is its volume, and $\chi_V (\bx)$ is the indicator function associated to the volume of space occupied by the rigid body around its center of mass. In the case of a homogeneous rigid sphere of radius $R \gg r_C$, $\Gamma^{\rm CSL} (\bD)$, $\Gamma^{\rm DP} (\bD)$, and $\Gamma^{\rm GPSL} (\bD)$ can be computed exactly (see Appendix~\ref{APPsec:DecoherenceRigidSphere}). The results are\footnote{The result for $\Gamma^{\rm DP} (\bD)$ is indeed equal to two times the one obtained by Penrose in Ref.~\cite{Penrose2014Gravitization} (Fig. 11).}
\begin{equations}\label{eq:RigidSphereDecoherenceTDCSL}
\Gamma^{\rm CSL} (\bD)
&=
\frac{3\gamma_{\rm CSL}}{4 \pi R^3}\frac{M^2}{m_0^2}\prtq{1-K^{\rm CSL}_{C}} 
+\\
&\qquad \qquad 
+\frac{R}{\gamma_{\rm CSL}} \frac{G^2 M^2 m_0^2}{\hbar^2} K^{\rm CSL}_{G} (\tl{D})
,\\
K^{\rm CSL}_{C} \prt{x}
&=\Theta(1-x)\prtq{1-\frac{3}{2}x+ \frac{x^3}{2}},\\
K^{\rm CSL}_{G} \prt{x}
&= \frac{\pi}{70}\prtqB{\Theta\prt{1-x}\prt{56 x^2 - 28 x^4 + 14 x^5 - x^7} 
	+\\ 
	&\qquad\qquad +\Theta\prt{x-1}\prt{70 x - 36 + 7/x}},
\end{equations}
for the CSL model,
\begin{equations}\label{eq:RigidSphereDecoherenceTDDP}
\Gamma^{\rm DP} (\bD) 
&=
2\frac{G M^2}{\hbar R} K^{\rm DP} (\tl{D}),\\
K^{\rm DP} \prt{x}
&=
\frac{\pi^{3/2}}{2\sqrt{2}}
\prtqB{\Theta\prt{1-x}\prt{4 x^2 - 3 x^3 + \frac{2}{5}x^5} 
	+\\ 
	&\qquad\qquad +\Theta\prt{x-1}\prt{\frac{12}{5}-\frac{1}{x}}},
\end{equations}
for the DP model, and
\begin{equations}\label{eq:RigidSphereDecoherenceGPSL}
\Gamma^{\rm GPSL} (\bD)
&= \gamma \frac{M}{m_0}\prtq{1-K_{C}^{\rm GPSL}(\tl{D})+ \frac{R_M^2}{R^2}K_{G}^{\rm GPSL} (\tl{D})},
\\
K_C^{\rm GPSL} (x) &= \Theta(1-x)\prtq{1-\frac{3}{2}x+ \frac{x^3}{2}},
\\
K_{G}^{\rm GPSL} \prt{x}
&=
\Theta (1-x) \frac{x^2}{2}\prtqB{\prt{3+60 x^2}\arccos(x)
	+\\
	&\qquad\qquad-x\sqrt{1-x^2}\prt{13 + 50 x^2}},
\end{equations}
for the GPSL model, where $\tl{D} = \abs{\bD}/(2R)$ and $R_M = (G m_0 M)(\gamma \hbar)^{-1}$. The functions $K_G (x)$ (plotted in Fig.~\ref{fig:plotDecoherenceRatesRigidSpheres}) behave quadratically for $\tl{D} \ll 1$ and reach unit magnitude for $\tl{D} \sim 1$. Because of the short ranged nature of the gravitational decoherence in the GPSL model, $K_{G}^{\rm GPSL} (\tl{D})$ vanishes for $\tl{D} \geq 1$. On the contrary, it saturates to a finite value in the DP model.
In all cases, when $\tl{D} \gg 1$ the decoherence rate dependence on the distance becomes the same as in the single particle case.

We introduced the length\footnote{We recall that, contrary to the single particle case of Eq.~\eqref{eq:SingleParticlePSLGravitationalDecoherence_Evaluated}, Eq.~\eqref{eq:RigidSphereDecoherenceGPSL} does not rely on an assumption such as $R_M \ll R$. The perturbative master equation of Eq.~\eqref{eq:PerturbativeGPSL_MasterEquation} relies on the mass density of the rigid sphere being not too high.}
$R_M$ so to clearly see what happens by considering spheres of the same density but different radius. Indeed, writing $M \sim \mu_0 R^3$ where, $\mu_0$ is the constant density of the sphere, we see that by considering larger and larger spheres, the gravitational decoherence term will eventually dominate the spontaneous collapse one (for $D\leq 1$). The scale at which the gravitational decoherence rate can be comparable with the spontaneous collapse one is around the human scale. For example, considering a body with a mass density of $100\ \rm{kg}\ \rm{m}^{-3}$, and using $\gamma=10^{-9} \rm{Hz}$ one has that $R_M \sim R$ when $R \sim \sqrt{\gamma \hbar/(G m_0 \mu_0)} \sim 10^{-4} \rm{m}$. The value $\gamma=10^{-9} \rm{Hz}$ is the lowest one allowed according the theoretical lower bound of Ref.~\cite{Piccione2025ExploringMassDependence}. Using the upper bound value of $\gamma$ when $r_C = 10^{-7} \rm{m}$, (i.e., $\gamma \sim 10^{-3} \rm{Hz}$) one would get $R\sim 10^{-1} \rm{m}$, which is still at the human scale. However, this is also the macroscopic scale at which objects behave classically. Therefore the most important term in the decoherence remains the spontaneous collapse one because it is linear in $D$ when $D\ll 1$, while the gravitational one is quadratic.

\section{Conclusions\label{Sec:Conclusions}}

We investigated the hybrid classical-quantum GPSL model presented in Ref.~\cite{Piccione2023Collapse}, which sources a classical Newtonian gravitational field from the random discrete collapses occurring on matter.
Compared to other hybrid models~\cite{Tilloy2016CSLGravity,Tilloy2017LeastDecoherence,GaonaReyes2021GravitationalFeedback}, GPSL does not allow for negative mass readings and makes the gravitational field associated to empty space \emph{exactly} vanishing. We argue that this makes the model more promising as the starting point for a relativistic hybrid model, since the vacuum needs to be stable in relativistic spontaneous collapse theories~\cite{Myrvold2017RelativisticMarkovian} to avoid intractable divergences.
In fact, we deem reasonable to assume that this requirement extends to any full hybrid classical-quantum model of gravity and should survive a non-relativistic limit.
Additionally, GPSL directly applies to indistinguishable particles and fields.

The GPSL model, with its clear and well-defined predictions, offers several avenues for experimental verification~\cite{Janse2024BoundsSpacetimeDiffusion}. Notably, its different mathematical structure leads to the avoidance of long-range gravitational decoherence and to the absence of gravitational field fluctuations in empty space, thus making it distinguishable from other hybrid gravity models. The model also allows, in principle, for lower decoherence than the bound set by the hybrid model based on the Diósi-Penrose spontaneous collapse model, thus enabling tests that could falsify all continuous weak measurement models without ruling out GPSL.

Additionally, the collapse parameters $\gamma$ and $r_C$, which govern the frequency and spatial resolution of the spontaneous measurements, are accessible to experimental bounds, particularly through decoherence measurements in spatial superposition experiments with mesoscopic or macroscopic objects~\cite{Carlesso2022Present}. Among the most promising routes are matter-wave interferometric tests with increasingly massive particles or rigid bodies~\cite{Arndt2014TestingSuperpositions}. At the same time, non-interferometric tests based on spontaneous radiation emission~\cite{Donadi2021NovelCSLBounds,Donadi2021UndergroundTest,MAJORANA2022WaveFunctionCollapse}, heating effects in bulk matter~\cite{Vinante2016BoundsNanoCantilevers,Vinante2017UltracoldCantilevers,Vinante2020UltracoldLayeredForce} or astronomical objects~\cite{Adler2019NeutroStarSpontaneousHeating}, or high-precision force measurements~\cite{Carlesso2016ExperimentalBounds,Helou2017LISACollapseModels} can provide independent constraints on the model's parameters and help identify or exclude the model’s predictions. These complementary strategies collectively position GPSL as a theoretically robust and experimentally approachable candidate among hybrid gravity models.

As a final comment, we remark that GPSL, as the other hybrid gravitational models in the Newtonian limit, should be seen as a theoretical first step in getting to a model merging quantum mechanics and general relativity. This task is obviously very challenging. An intermediate step could consist of exploring the first-order post-Newtonian approximations to the model. This has already been done for the Schr\"{o}dinger-Newton equation, e.g. in Refs.~\cite{Manfredi2015SchrodingerNewtonBeyondNewton,Brizuela2022PostNewtonianSchrNewt}; the same kind of modifications should be applicable to GPSL without losing its advantages over the Schr\"{o}dinger-Newton equation.

\section*{Acknowledgements}

N.P. thanks J. L. Gaona-Reyes for useful explanation about the Tilloy-Diósi prescription. 
N. P. thanks  J. L. Gaona-Reyes, O. Angeli, and D. G. A. Altamura for useful discussions about the results of this work. 
N.P thanks M. Carlesso for useful comments about the Lindblad form of the perturbed GPSL master equation. 
N.P thanks A. Gundhi for useful discussions about the contributions of the gravitational feedback to the decoherence of the GPSL model and decoherence in general.
N. P. thanks O. Angeli for useful discussions about relevant literature related to this work.
We thank A. Tilloy and L. Diósi for useful comments about this work.

The authors acknowledge support from the PNRR MUR projects PE0000023-NQSTI, INFN and the University of Trieste.
N. P. acknowledges support also from the European Union Horizon’s 2023 research and innovation programme [HORIZON-MSCA-2023-PF-01] under the Marie Sklodowska Curie Grant Agreement No. 101150889 (CPQM).
A. B. acknowledges support also from the EU EIC Pathfinder project QuCoM (Grant Agreement No. 101046973).


%

\onecolumngrid
\appendix

\clearpage
\section{The Tilloy-Diósi Models\label{APPsec:TDmodelGeneral}}

In this appendix, we will resume the continuous weak monitoring models following Refs.~\cite{Tilloy2016CSLGravity,Tilloy2017LeastDecoherence,GaonaReyes2021GravitationalFeedback}. These models are constructed by considering naturally occurring continuous weak measurements of the smeared mass density\footnote{We recall that the smeared mass density operator is defined by $\hmu_{r_C} (\bx) = (g_{r_C} * \hmu) (\bx)$ where $g_{r_C} (\bx)$ is a smearing function usually characterized by a radius $r_C$ and the $*$ operator denotes convolution. Usually, one takes the Gaussian smearing $g_{r_C}(\bx) = (2 \pi r_C^2)^{-3/2}\exp{-\bx^2/(2 r_C^2)}$. This smearing is necessary to avoid divergences and $r_C$ usually constitutes a free parameter of spontaneous collapse models.} $\hmu_{r_C} (\bx)$ at all points in space, and implementing Newtonian gravity as a feedback mechanism based on the measurements' results. For future convenience, we also introduce the notation $\tmuR (\bx) = \hmu_{r_C} (\bx) - \ev{\hmu_{r_C} (\bx)}$.

\subsection{The Measurement Part\label{subsec:TD_model_MeasurementPart}}

We consider that at each spatial point a continuous weak measurement of the smeared density mass operator takes place, giving the measurement record~\cite{GaonaReyes2021GravitationalFeedback}:
\begin{equation}
\mu_t (\bx) = \ev{\hmu_{r_C} (\bx)} + \delta \mu_t (\bx),
\qquad
\delta \mu_t (\bx) = \frac{1}{2}\int \dd[3]{\by} \gamma_C^{-1} (\bx-\by)\frac{\dd{W_t} (\by)}{\dd{t}},
\end{equation}
where the Wiener increment $\dd{W_t} (\bx)$ is such that\footnote{This kind of generalized Wiener increment can be obtained by convoluting the usual Wiener increment. See page 22 (492) of Ref.~\cite{Bassi2013Models}.} $\dd{W_t} (\bx) \dd{W_t} (\by) = \gamma_C (\bx,\by) \dd{t} = \gamma_C \prt{\abs{\bx-\by}} \dd{t}$. 
Moreover, one has $\dd{W_t} (\bx) \dd{t} =0$ and $\mbE [\dd{W_t} (\bx)] = 0$.
The inverse of $\gamma_C (\bx,\by)$ is defined by the relation 
\begin{equation}
(\gamma_C \circ \gamma_C^{-1} )(\bx,\by) \equiv 
\int \dd[3]{\bz} \gamma_C (\bx,\bz)\gamma_C^{-1} (\bz,\by) =
\int \dd[3]{\bz} \gamma_C (\bx-\bz)\gamma_C^{-1} (\bz-\by) = \delta (\bx-\by).
\end{equation}
This leads to the stochastic equation
\begin{equation}
\prtq{\dd{\ket{\psi_t}}}_{\rm meas}
=
\prtq{\int \dd[3]{\bx} \tmuR (\bx) \dd{W_t} (\bx) - \frac{1}{2}\int \dd[3]{\bx}\dd[3]{\by} \gamma_C (\bx,\by) \tmuR(\bx)\tmuR(\by) \dd{t}}\ket{\psi_t},
\end{equation}
where $\prtq{\dd{\ket{\psi_t}}}_{\rm meas}$ denotes the infinitesimal variation due to the measurements. Notice how $\gamma_C (\bx,\by)$ enters in the stochastic equation while the measurement error is defined through $\gamma_C^{-1} (\bx,\by)$. This can be intuitively understood because the less information one gets from the measurement ($\gamma_C^{-1} (\bx,\by)$ is somewhat \enquote{large}) the less the measured system is disturbed ($\gamma_C (\bx,\by)$ is somewhat \enquote{small}). 

The master equation can be obtained by averaging over the stochastic equation for the density matrix: $\dd{\rho_t} = \mbE \prtq{\dd{\sigma_t}}= \mbE \prtq{\dyad{\dd{\psi_t}}{\psi_t} + \dyad{\psi_t}{\dd{\psi_t}} + \dyad{\dd{\psi_t}}}$. One gets
\begin{equation}\label{APPeq:TDModelsMasterEquationMeasurementPart}
\dv{t} \rho_t = - \frac{1}{2}\int \dd[3]{\bx}\dd[3]{\by} \gamma_C (\bx,\by) \comm{\hmu_{r_C} (\bx)}{\comm{\hmu_{r_C} (\by)}{\rho_t}}.
\end{equation}
The CSL model is recovered by choosing $\gamma_C (\bx,\by) = (\gamma_{\rm CSL}/m_0^2) \delta (\bx-\by)$ and making the substitution $\dd{W_t} \rightarrow (\sqrt{\gamma_{\rm CSL}}/m_0) \dd{W_t}$. The DP model is instead obtained by choosing $\gamma_C (\bx,\by) = -(1/2\hbar)\mcV(\bx,\by)$, where we recall that $\mcV(\bx,\by)=-G\abs{\bx-\by}^{-1}$.

\subsection{Adding the Gravitational Feedback}

Newtonian gravity can be obtained by introducing the classical potential
\begin{equation}
\Phi_{C} (\bx,t) = \int \dd[3]{\by} \mcV(\bx-\by)\mu_{t} (\by),
\end{equation}
where $\mcV(\bx-\by)= - G \abs{\bx-\by}^{-1}$ and $G$ is Newton's constant. We immediately see how taking the average $\mbE \prtq{\Phi_{C} (\bx,t)}$ gives the usual semiclassical Newtonian potential when considering distances much higher than $r_C$. 

The procedure followed in Ref.~\cite{Tilloy2016CSLGravity} is akin to a measurement and feedback procedure~\cite{Book_Wiseman2009Measurement,Book_Jacobs2014MeasurementTheory} with a detector performing a continuous weak measurement at each spatial point. The feedback Hamiltonian chosen in Ref.~\cite{Tilloy2016CSLGravity,GaonaReyes2021GravitationalFeedback} is 
\begin{equation}
\hat{H}_{\rm fb} (t) 
= \int \dd[3]{\bx} \Phi_{C} (\bx,t) \hmu_{r_C} (\bx)
= \int \dd[3]{\bx} \dd[3]{\by} \mcV(\bx-\by) \mu_t (\by) \hmu_{r_C} (\bx),
\end{equation}
where we remark that the mass operator appearing in $H_{\rm fb} (t)$ is also smeared. Implementing the feedback one gets the following three contributions to the differential of the wavefunction:
\begin{equations}\label{eq:TDFeedbackSchrodingerEquation}
\prtq{\dd{\ket{\psi_t}}}_{\rm meas}
&=
\prtq{\int \dd[3]{\bx} \tmuR (\bx) \dd{W_t} (\bx) - \frac{1}{2}\int \dd[3]{\bx}\dd[3]{\by} \gamma_C (\bx,\by) \tmuR(\bx)\tmuR(\by) \dd{t}}\ket{\psi_t},\\
\prtq{\dd{\ket{\psi_t}}}_{\rm fb} &= \prtqB{
	-\frac{i}{\hbar} \int \dd[3]{\bx}\dd[3]{\by} \mcV (\bx-\by) \ev{\hmu_{r_C} (\by)} \hmu_{r_C} (\bx) \dd{t}
	- \frac{i}{2\hbar}\int \dd[3]{\bx}\dd[3]{\by} (\mcV \circ \gamma_C^{-1}) (\bx-\by) \hmu_{r_C} (\bx) \dd{W_t} (\by) +\\
	&\qquad 
	- \frac{1}{8 \hbar^2} \int \dd[3]{\bx}\dd[3]{\by} (\mcV \circ \gamma_C^{-1} \circ \mcV) (\bx-\by) \hmu_{r_C} (\bx) \hmu_{r_C} (\by)\dd{t}}\ket{\psi_t},\\
\prtq{\dd{\ket{\psi_t}}}_{\rm corr} &=  \prtq{-\frac{i}{2\hbar} \int \dd[3]{\bx}\dd[3]{\by} \mcV(\bx-\by) \tmuR (\by) \hmu_{r_C} (\bx)\dd{t}}\ket{\psi_t}.
\end{equations}
From the above equations, the master equation can be obtained by averaging over the stochastic master equation. The result is
\begin{equation}\label{APPeq:GravitationalMasterEquationTDModels}
\dv{t} \rho_t =  
-\frac{i}{\hbar}\comm{\int \dd[3]{\bx}\dd[3]{\by} \frac{\mcV (\bx-\by)}{2} \hmu_{r_C} (\bx) \hmu_{r_C} (\by)}{\rho_t} 
- \frac{1}{2} \int \dd[3]{\bx}\dd[3]{\by} \mcD_{\rm CM} (\bx,\by) \comm{\hmu_{r_C} (\bx)}{\comm{\hmu_{r_C} (\by)}{\rho_t}},
\end{equation}
where
\begin{equation}
\mcD_{\rm CM} (\bx,\by) = \gamma_C (\bx,\by) + \frac{1}{4 \hbar^2}(\mcV \circ \gamma_C^{-1} \circ \mcV) (\bx,\by).
\end{equation}
Comparing Eq.~\eqref{APPeq:GravitationalMasterEquationTDModels} with Eq.~\eqref{APPeq:TDModelsMasterEquationMeasurementPart} one sees that an Hamiltonian term appears due to the feedback. This term is exactly equal to the standard quantization of the Newtonian potential when $r_C \rightarrow 0$. Thus, in the continuous weak monitoring approach, this term is responsible for accounting Newtonian gravitation but it also predicts modifications of it at lengthscales lower or similar to $r_C$.

\subsection{Stochastic Unraveling of Continuous weak Monitoring Models Master Equations\label{APPSubsec:UnitaryUnravelingTDModels}}

It usually convenient to unravel the master equation of a spontaneous collapse model in terms of a stochastic potential. This is a standard procedure since Ref.~\cite{Fu1997SpontaneousRadiation}.

Let us consider a master equation of the following kind:
\begin{equation}\label{APPeq:GeneralHamiltonianTDMasterEquation}
\dv{t} \rho_t =  
-\frac{i}{\hbar}\comm{\hat{H}}{\rho_t} 
- \frac{1}{2} \int \dd[3]{\bx}\dd[3]{\by} \mcD_{\rm CM} (\bx,\by) \comm{\hL (\bx)}{\comm{\hL (\by)}{\rho_t}},
\end{equation}
We want to prove that there exist a unitary unraveling given by the following stochastic potential:
\begin{equation}
\hV(t) = \hbar \int \dd[3]{\bx} \hL (\bx) w(\bx,t),
\end{equation}
where $w(\bx,t)$ is a white-in-time noise characterized by 
\begin{gather}
\mbE\prtq{w(\bx,t)}=0,
\qquad
\mbE \prtq{w(\bx,t)w(\by,s)}=\delta\prt{t-s} \mcD_{\rm CM}(\bx,\by),
\end{gather}
where $\mcD_{\rm CM}(\bx,\by) = \mcD_{\rm CM} (\abs{\bx-\by})$.

To prove the validity of the unraveling one starts with a density matrix $\rho_t$ and evolves in time to $t+\delta t$, up to second order in $\delta t$, according to a certain realization of the noise with
\begin{equation}
U(t+\delta t,t) = \mcT \exp{-\frac{i}{\hbar}\int_t^{t+\delta t} H'(s) \dd{s}}
\simeq 1 - \frac{i}{\hbar}\int_t^{t+\delta t} H'(s) \dd{s} -\frac{1}{\hbar^2}\int_t^{t+\delta t}\int_t^{t_1}  H'(t_1)H'(t_2) \dd{t_1}\dd{t_2},
\end{equation}
where $H'(t) = \hat{H} + \hV (t)$.
The average density matrix is obtained by taking the average over the noise of $\rho_{t+\delta t} = U(t+\delta t,t)\rho_t U^\dg(t+\delta t,t)$ and keeping only terms up to first order in $\delta t$. The heuristic argument\footnote{This connected to the idea of writing $w(\bx,t) = \dd{W_t (\bx)}/\dd{t}$, where $\dd{W_t} (\bx)$ is the generalized Wiener increment.} to do so is to consider $\prtq{w(\bx,t)}\sim \delta t^{-1/2}$ so that, for example, a term like $\int \hat{H} \hV (t_2)\dd{t_1}\dd{t_2}$ and $\int \hV (t_1)\hV(t_2)\hV(t_3) \dd{t_1}\dd{t_2}\dd{t_3}$ are both of higher order and have to be neglected. Then, one must use equalities such as $\mbE \prtq{\int_t^{t+\delta t} \hV(s)\dd{s}}=0$,
\begin{equation}
\mbE\prtq{\int_t^{t+\delta t}\int_t^{t+\delta t} \dd{t_1} \dd{t_2} \hV(t_1)\rho_t \hV (t_2)}
= \hbar^2 \delta t \int \dd[3]{\bx}\dd[3]{\by} \mcD_{\rm CM}(\bx,\by) \hL(\bx)\rho_t \hL(\by),
\end{equation}
and
\begin{multline}
\mbE\prtq{\int_t^{t+\delta t}\int_t^{t_1} \dd{t_1} \dd{t_2} \hV(t_1)\hV(t_2)\rho_t}
= \hbar^2 \prtq{\int_t^{t+\delta t}\int_t^{t_1} \delta(t_1-t_2)\dd{t_1}\dd{t_2}} \int \dd[3]{\bx}\dd[3]{\by} \mcD_{\rm CM}(\bx,\by)\hL(\bx)\hL (\by)\rho_t
=\\=
\frac{\delta t}{2}\hbar^2 \int \dd[3]{\bx}\dd[3]{\by} \mcD_{\rm CM}(\bx,\by)\hL(\bx)\hL (\by)\rho_t.
\end{multline}
The master equation is finally obtained by writing $\dot{\rho}_t = \lim_{\delta t \rightarrow 0} (\rho_{t+\delta t} - \rho_t)/\delta t$.

Now that we have proven that Eq.~\eqref{APPeq:GeneralHamiltonianTDMasterEquation} can be unraveled by
\begin{equation}
i\hbar \dv{t} \ket{\psi_t} = \prtq{\hat{H} + \hV (t)}\ket{\psi_t},
\end{equation}
it is easy to show that the average force acting on each particle and due to $\hV (t)$ is vanishing on average. If $\hat{H}$ is the pairwise gravitational interaction (with or without smearing), conservation of average momentum follows automatically.

\subsection{Decoherence Rate for Isolated Particles\label{APPSubsec:TDModelSingleParticlesDecoherenceRate}}

In a generic continuous weak monitoring model, the single particle decoherence takes the form
\begin{equations}
\dv{t} \rho_t (\bx,\by) 
&= - m^2 \rho_t (\bx,\by) \int \dd[3]{\bk} \mcF \prtg{\mcD_{\rm CM}} (\bk) \prtq{\mcF \prtg{g_{r_C}} (\bk)}^2 \prtq{1 - \cos(\bk\cdot(\bx-\by))},\\
&= - 4 \pi m^2 \rho_t (\bx,\by) \int_0^{\infty} \dd{k} k^2 \mcF \prtg{\mcD_{\rm CM}} (k) \frac{e^{-k^2 r_C^2}}{(2\pi)^{3}} \prtq{1 - \frac{\sin(k \abs{\bx-\by})}{k \abs{\bx-\by}}},
\end{equations}
where $\mcF \prtg{f} (\bk)$ denotes the Fourier transform of $f(\bx)$. In the above equation, $\mcF \prtg{\mcD_{\rm CM}} (\bk)$ is an abuse of notation due to the fact that $\mcD_{\rm CM}(\bx,\by)$ is translation invariant and, therefore, its Fourier transform is diagonal in momentum space. Using the above equation one can obtain the following decoherence rate for the continuous weak monitoring model based on the DP spontaneous collapse model:
\begin{equation}
\mcF\prtg{\mcD_{\rm DP}}(\bk) = \frac{4 G \pi}{\hbar \bk^2}
\implies
\Gamma^{\rm DP} (\bx,\by) = \frac{G m^2}{\hbar r_C \sqrt{\pi}}\prtq{1 - \frac{\sqrt{\pi}}{2\td}\erf\prt{\td}},
\end{equation}
where we introduced $\td \equiv \abs{\bx-\by}/(2 r_C)$. 

The decoherence rate of the continuous weak monitoring model based on the CSL spontaneous collapse model can be obtained by summing the collapse part (see, for example, Ref.~\cite{Piccione2025ExploringMassDependence}) and the gravitational part, which is equal to
\begin{equation}
\Gamma_{G}^{\rm CSL} (\bD) = 
\frac{m_0^2 m^2 G^2}{8 \hbar^2 \gamma_{\rm CSL}} \int \dd[3]{\bz} \prtq{\int \dd[3]{\bz'}\frac{1}{\abs{\bz-\bz'}}\prtq{g_{r_C} (\bz') - g_{r_C} (\bz'+\bD)}}^2.
\end{equation}
If we take the limit $r_C \rightarrow 0 \implies g_{r_C} (\bx) \rightarrow \delta (\bx)$ one can exploit the equality
\begin{equation}
\int \dd[3]{\bz} \prtq{\frac{1}{\abs{\bz}}-\frac{1}{\abs{\bz+\bD}}}^2
=
4 \pi \abs{\bD}.
\end{equation}
to get the same formula of Eq.~(33) [page 6] of Ref.~\cite{Tilloy2016CSLGravity}, even though with some different notational choices. Keeping $r_C$ finite, instead, the calculation proceeds as follows. First, since the result cannot depend on the direction of $\bD$, we set $\bD = d \hat{e}_3$ and rewrite the integral as
\begin{equation}
\int \dd[3]{\bz}\dd[3]{\bz'}\dd[3]{\bz''} G_2(\bz',d)G_2(\bz'',d)\frac{1}{\abs{\bz-\bz'}}\frac{1}{\abs{\bz-\bz''}},
\end{equation}
where $G_2(\bz',d) = \prtq{g_{r_C} (\bz')-g_{r_C} (\bz'-d\hat{e}_3)}$. This integral can then be rewritten as follows, remembering that $\mcF \prt{\abs{\bz-\bz'}^{-1}} (\bk,\bk') = 4\pi \bk^{-2} \delta (\bk+\bk')$:
\begin{equation}
\int \dd[3]{\bz}\dd[3]{\bz'}\dd[3]{\bz''} G_2(\bz',d)G_2(\bz'',d)\frac{1}{\abs{\bz-\bz'}}\frac{1}{\abs{\bz-\bz''}}
=
16 \pi^2 \int \dd[3]{\bk} 
\frac{\tl{G}_2 (\bk,d)\tl{G}_2 (-\bk,d)}{\bk^4}.
\end{equation}
The Fourier transform of $G_2$ is given by
\begin{equation}
\tl{G}_2 (\bk,d) = \frac{1-e^{i d k_3}}{(2\pi)^{3/2}}e^{-\bk^2 r_C^2/2}
\implies
\tl{G}_2 (\bk,d)\tl{G}_2 (-\bk,d) = \frac{2 e^{- \bk^2 r_C^2}}{(2 \pi)^3} \prtq{1 - \cos(d k_3)}.
\end{equation}
So, the integral becomes
\begin{equation}
\frac{32 \pi^2}{(2\pi)^3} \int \dd[3]{\bk} 
\frac{\frac{e^{- \bk^2 r_C^2}}{(2 \pi)^3} \prtq{1 - \cos(d k_3)}}{\bk^4}
=
8\sqrt{\pi}r_C \prtq{\sqrt{\pi} \prt{\td +\frac{1}{2\td}}\erf(\td) -\prt{2-e^{-\td^2}}},
\end{equation}
where in the last line we used $\td =d/(2 r_C)$. We also have that
\begin{equations}
\td \ll 1 &\implies 8\sqrt{\pi}r_C \prtq{\sqrt{\pi} \prt{\td +\frac{1}{2\td}}\erf(\td) -\prt{2-e^{-\td^2}}} \simeq \frac{16 \sqrt{\pi}}{3} r_C \td^2,\\
\td \gg 1 &\implies 8\sqrt{\pi}r_C \prtq{\sqrt{\pi} \prt{\td +\frac{1}{2\td}}\erf(\td) -\prt{2-e^{-\td^2}}} \simeq 4 \pi r_C \prt{2 \td - \frac{4}{\sqrt{\pi}} + \frac{1}{\td}} \simeq 8 \pi r_C \td.
\end{equations}
In conclusion, the decoherence rate due to gravity is
\begin{equation}
\Gamma_{G}^{\rm CSL} (\bD) = 
\frac{\pi r_C}{\gamma_{\rm CSL}} \frac{m_0^2 m^2 G^2}{\hbar^2}  \prtq{\prt{\td +\frac{1}{2\td}}\erf(\td) -\frac{2-e^{-\td^2}}{\sqrt{\pi}}}
\end{equation}

\clearpage
\section{Field Fluctuations\label{APPSec:FieldFluctuations}}

In this appendix, we compute the fluctuations for both the GPSL and continuous weak monitoring models.
We start with the GPSL model, recalling that
\begin{equation}
\Phi (\bx,t)\dd{t} = \frac{m_0}{\gamma}\int \dd[3]{\by} \mcV(\bx,\by) \dd{N_t} (\by), 
\quad
\mbE \prtq{\dd{N_t} (\by)} = \frac{\gamma}{m_0}\ev{\hmu_{r_C} (\by)}_t\dd{t},
\quad
\mbE \prtq{\Phi (\bx,t)\dd{t}} = \int \dd[3]{\by} \mcV(\bx,\by)\ev{\hmu_{r_C} (\by)}_t \dd{t}
\end{equation}
which implies that
\begin{equation}
\mbE \prtq{\Phi (\bx,t)\dd{t}\Phi (\bx',t)\dd{t}}
= \frac{m_0}{\gamma} \int \dd[3]{\by}\mcV(\bx,\by)\mcV(\bx',\by) \ev{\hmu_{r_C} (\by)}_t\dd{t}.
\end{equation}
Since the term $\mbE \prtq{\Phi (\bx,t)\dd{t}}\mbE \prtq{\Phi (\bx',t)\dd{t}}$ is proportional to $\dd{t}^2$, we can neglect it and the above expression is also the covariance of the Newtonian field sourced by the PSL model. Its square root gives the fluctuations of the field. The variance is given by the covariance by setting $\bx=\bx'$:
\begin{equation}
\mbE \prtq{\Phi (\bx,t)\dd{t}\Phi (\bx,t)\dd{t}}
= \frac{m_0}{\gamma} \int \dd[3]{\by}\mcV^2(\bx,\by)\ev{\hmu_{r_C} (\by)}_t\dd{t}
= \frac{m_0 G^2}{\gamma} \int \dd[3]{\by}\frac{\ev{\hmu_{r_C} (\bx+\by)}_t}{\by^2}\dd{t}.
\end{equation}

Regarding the continuous weak monitoring case, we recall that:
\begin{equation}
\mu_t (\bx) = \ev{\hmu_{r_C} (\bx)} + \delta \mu_t (\bx),
\qquad
\delta \mu_t (\bx) = \frac{1}{2}\int \dd[3]{\by} \gamma_C^{-1} (\bx-\by)\frac{\dd{W_t} (\by)}{\dd{t}},
\end{equation}
where the Wiener increment $\dd{W_t} (\bx)$ is such that $\dd{W_t} (\bx) \dd{W_t} (\by) = \gamma_C (\bx,\by) \dd{t} = \gamma_C \prt{\abs{\bx-\by}} \dd{t}$ and the inverse of $\gamma_C (\bx,\by)$ is defined by the relation 
\begin{equation}
(\gamma_C \circ \gamma_C^{-1} )(\bx,\by) \equiv 
\int \dd[3]{\bz} \gamma_C (\bx,\bz)\gamma_C^{-1} (\bz,\by) =
\int \dd[3]{\bz} \gamma_C (\bx-\bz)\gamma_C^{-1} (\bz-\by) = \delta (\bx-\by),
\end{equation}
and
\begin{equation}
\Phi_{C} (\bx,t)\dd{t} = \int \dd[3]{\by} \mcV(\bx,\by)\mu_{t} (\by)\dd{t}.
\end{equation}
Then, one has
\begin{equation}
\mbE \prtq{\delta \mu_t (\bx)\delta \mu_t (\by) \dd{t}^2}
=
\frac{\gamma_C^{-1} (\bx,\by)}{4}\dd{t}
\implies
\mbE \prtq{\Phi_C (\bx,t)\dd{t}\Phi_C (\bx',t)\dd{t}}
= \frac{1}{4} \int \dd[3]{\by}\dd[3]{\by'} \mcV(\bx,\by)\mcV(\bx',\by') \gamma_C^{-1} (\by,\by')\dd{t},
\end{equation}
which can be written as
\begin{equation}
\mbE \prtq{\Phi_C (\bx,t)\dd{t}\Phi_C (\bx',t)\dd{t}}
=
\frac{\dd{t}}{4}(\mcV \circ \gamma_C^{-1} \circ \mcV)(\bx,\bx').
\end{equation}
So, focusing on the CSL and DP cases:
\begin{equations}
\textrm{CSL:}
\quad
\gamma_C (\bx,\by) = \frac{\gamma_{\rm CSL}}{m_0^2}\delta (\bx-\by)
&\implies
\mbE \prtq{\Phi_C (\bx,t)\dd{t}\Phi_C (\bx',t)\dd{t}}
=
\frac{\dd{t}}{4}\frac{m_0^2}{\gamma_{\rm CSL}}(\mcV \circ \mcV)(\bx,\bx')
\\
\textrm{DP:}
\quad
\gamma_C (\bx,\by) = -\frac{1}{2\hbar} \mcV(\bx,\by)
&\implies
\mbE \prtq{\Phi_C (\bx,t)\dd{t}\Phi_C (\bx',t)\dd{t}}
=
-\frac{\hbar \dd{t}}{2}\mcV(\bx,\bx').
\end{equations}
In both cases, the variance (the covariance with $\bx=\bx'$) diverges.

\clearpage
\section{Conservation of Average Momentum and Pairwise Average Forces\label{APPsec:ConservationAverageMomentum}}

In this appendix, we prove that the total average momentum of an isolated system governed only by the GPSL dynamics is conserved. 

Let us consider an ensemble of $N$ particles. 
Since the processes of spontaneous collapse and gravitational back-reaction are instantaneous and tied together in the stochastic and master equations [Eq.~\eqref{eq:PSLDynamicsWithGravityParticles}], it is not as easy as one might expect from similar analysis carried out for other models such as the continuous weak monitoring models, to show that the gravitational feedback contribution does not change on average the total momentum. To face this problem, we smear the gravitational feedback in time by an amount $\tau$ so to formally separate it from the collapse process. One then makes the limit $\tau \rightarrow 0$ at the end of the calculations to restore the original dynamics. We can write the stochastic equation of Eq.~\eqref{eq:PSLDynamicsWithGravityParticles} of the main text as follows:
\begin{equation}\label{APPeq:NonMarkovianGPSL}
\dd{\ket{\psi_t}}
= \prtq{-\frac{i}{\hbar} \prt{\hat{H}+\frac{1}{\gamma}\int_{t-\tau}^{t}\dd{s} f_\tau (t-s) \int \dd[3]{\bx} \hV (\bx) I(s,\bx)}\dd{t} + \int \dd[3]{\bx} \prt{\frac{\ghm (\bx)}{\sqrt{\ev{\hmu_{r_C} (\bx)}}}-1}\dd{N}_t (\bx)} \ket{\psi_t},
\end{equation}
where $f_\tau (t)$ is a smearing function different from zero only in a small interval of length $\tau$ and $I(t,\bx) = \dd{N_t} (\bx)/\dd{t}$ is the so-called measurement current (see page 256 of Ref.~\cite{Book_Wiseman2009Measurement}). In particular, we assume that $f_\tau (0)=f_\tau (\tau)=0$ and $\int_{0}^{\tau} f_\tau (s) \dd{s} =1$. Moreover, we recall that
\begin{equation}
\hV (\bx) = - G m_0 \sum_k m_k \int \dd[3]{\by} \frac{1}{\abs{\bx-\by}}g_{r_C} (\by-\hbq_k) = - G m_0 \sum_k m_k \frac{1}{\abs{\bx-\hbq_k}}\erf \prt{\frac{\abs{\bx-\hbq_k}}{r_C\sqrt{2}}}.
\end{equation}
To check that Eq.~\eqref{APPeq:NonMarkovianGPSL} gives back Eq.~\eqref{eq:PSLDynamicsWithGravityParticles} when $\tau \rightarrow 0$ let us consider what happens after a collapse at point $\bx_c$ and time $t_c$ so that $I (t,\bx)=\delta(t-t_c) \delta(\bx-\bx_c)$. Since $\tau \rightarrow 0$, we can safely assume that no other collapses take place for $t \in [t_c,t_c + \tau]$ and that we can neglect the standard Hamiltonian of the system. Then, the state of the system at time $t_c+\tau$ is 
\begin{multline}
\lim_{\tau \rightarrow 0}\ket{\psi_{t_c+\tau}} =
\lim_{\tau \rightarrow 0}\exp{-\frac{i}{\hbar}\int_{t_c}^{t_c+\tau}\dd{t}
	\prtq{\frac{1}{\gamma}\int_{t-\tau}^{t}\dd{s} f_\tau (t-s) \int \dd[3]{\bx} \hV (\bx) \delta(s-t_c) \delta(\bx-\bx_c) }}
\frac{\ghm (\bx_c)}{\sqrt{\ev{\hmu_{r_C} (\bx_c)}_{t_c^-}}}
\ket{\psi_{t_c^-}}
=\\=
\lim_{\tau \rightarrow 0}\exp{-\frac{i}{\hbar \gamma}\int_{t_c}^{t_c+\tau}\dd{t}
	\prtq{f_\tau (t-t_c) \hV (\bx_c)}}
\frac{\ghm (\bx_c)}{\sqrt{\ev{\hmu_{r_C} (\bx_c)}_{t_c^-}}}
\ket{\psi_{t_c^-}}
=
e^{-(i/\hbar \gamma) \hV (\bx_c)}
\frac{\ghm (\bx_c)}{\sqrt{\ev{\hmu_{r_C} (\bx_c)}_{t_c^-}}}
\ket{\psi_{t_c^-}},
\end{multline}
where $t_c^-$ denotes a time infinitesimally before $t_c$ so that $\ket*{\psi_{t_c^-}}$ is the state before the collapse event. The state at the end of the above chain of equalities is exactly the state one would get from the stochastic equation of Eq.~\eqref{eq:PSLDynamicsWithGravityParticles} after a collapse.

Continuing with the assumption that a collapse takes place at time $t_c$ and point $\bx_c$, we can compute the (stochastic) force $\hF_{k,d} (t,\bx_c)$ acting on the $k$-th particle along an arbitrary direction $d$ and due to the gravitational feedback at a time $t$ such that $t_c< t < t_c+\tau$. Since, in the end, we take $\tau \rightarrow 0$, we can again assume that no other collapse happens and we can neglect the standard Hamiltonian, thus obtaining:
\begin{multline}
\hF_{k,d} (t,\bx_c)
= \frac{i}{\hbar}\comm{\frac{1}{\gamma}\int_{t_c}^{t} f_\tau (t-s) \int \dd[3]{\bx} \hV (\bx) \delta(s-t_c) \delta(\bx-\bx_c)\dd{s}}{\hp_{k,d}}
=\\=
\frac{i}{\hbar \gamma}f_\tau (t-t_c)\comm{ \hV (\bx_c)}{\hp_{k,d}}
=
-\frac{f_\tau (t-t_c)}{\gamma} \partial_{d}\prtq{\frac{G m_0 m_k}{\abs{\bx_c-\hbq_k}}\erf \prt{\frac{\abs{\bx_c-\hbq_k}}{r_C\sqrt{2}}} }
=\\=
\frac{G m_k m_0}{\gamma} f_\tau (t-t_c) \frac{(\bx_c)_d - \hq_{k,d}}{\abs{\bx_c-\hbq_k}^2}
\prtq{\frac{1}{\abs{\bx_c-\hbq_k}}\erf \prt{\frac{\abs{\bx_c-\hbq_k}}{r_C\sqrt{2}}}- 4 \pi r_C^2 g_{r_C} (\bx_c-\hbq_k)},
\end{multline}
which means that the impulse given to the $k$-th particle along the direction $d$ is
\begin{equation}
\hat{J}_{k,d} (\bx_c) 
= \int_{t_c}^{t_c + \tau} \hF_{k,d} (t,\bx_c) \dd{t}
= \frac{G m_k m_0}{\gamma} \frac{(\bx_c)_d - \hq_{k,d}}{\abs{\bx_c-\hbq_k}^2}\prtq{\frac{1}{\abs{\bx_c-\hbq_k}}\erf \prt{\frac{\abs{\bx_c-\hbq_k}}{r_C\sqrt{2}}}- 4 \pi r_C^2 g_{r_C} (\bx_c-\hbq_k)}.
\end{equation}
Since the collapse probability is proportional to the mass density [cf. Eq.~\eqref{eq:PoissonProcessCharacterization}], the conditional probability for a collapse to happen at $\bx_c \in \dd[3]{\bx}$ given that it happens at time $t_c$ is
\begin{equation}
P\prt{\bx_c \in \dd[3]{\bx}\vert \textrm{Collapse at time}\ t_c} = \frac{\ev{\hmu_{r_C} (\bx_c)}_{t_c^-} \dd[3]{\bx}}{\int \dd[3]{\bz} \ev{\hmu_{r_C} (\bz)}_{t_c^-}}= \frac{1}{M}\ev{\hmu_{r_C} (\bx_c)}_{t_c^-} \dd[3]{\bx},
\end{equation}
which implies that the average impulse is
\begin{equation}
\mbE \prtq{\hat{J}_{k,d} | \textrm{Collapse at time}\ t_c}
=
\frac{1}{M}\int \dd[3]{\bx} \ev{\hmu_{r_C} (\bx)}_{t_c^-} \hat{J}_{k,d} (\bx) \dd[3]{\bx}.
\end{equation}
The next step is to take also the average of the above quantity over the quantum state; to this end, we have to keep in mind that the state at time $t_c^+$ (a time infinitesimally after $t_c$ but previous to the gravitational feedback) on which the impulse acts depends on where the collapse happened. This state is given by
\begin{equation}
\ket{\psi_{t_c^+}} = \frac{\ghm (\bx_c)}{\sqrt{\ev{\hmu_{r_C} (\bx)}_{t_c^-}}}
\ket{\psi_{t_c^-}}.
\end{equation}
Therefore, the average of $\hat{J}_{k,d} (\bx)$ on both the noise and the state is given by
\begin{multline}\label{APPeq:AveragedImpulseCollapse}
\mbE \prtq{\ev{\hat{J}_{k,d}}_{t_c^+}| \textrm{Collapse at time}\ t_c}
=
\frac{1}{M} \int \dd[3]{\bx} \ev{\hmu_{r_C} (\bx)}_{t_c^-} 
\ev{\frac{\ghm (\bx)}{\sqrt{\ev{\hmu_{r_C} (\bx)}_{t_c^-}}}
	\hat{J}_{k,d} (\bx)
	\frac{\ghm (\bx)}{\sqrt{\ev{\hmu_{r_C} (\bx)}_{t_c^-}}}}_{t_c^-}
=\\=
\frac{1}{M} \int \dd[3]{\bx} \ev{\hmu_{r_C} (\bx) \hat{J}_{k,d} (\bx)}_{t_c^-}
=
\sum_j \frac{m_j}{M} \int \dd[3]{\bx} \ev{g_{r_C} (\hbq_j-\bx) \hat{J}_{k,d} (\bx)}_{t_c^-},
\end{multline}
where we exploited the fact that $\comm{\ghm (\bx)}{\hat{J}_{k,d} (\bx)}=0$ and we have used $\hmu_{r_C} (\bx) = \sum_j m_j g_{r_C} (\hbq_j-\bx)$. 

Each addendum in Eq.~\eqref{APPeq:AveragedImpulseCollapse} can be seen as the average impulse applied by the $j$-th particle on the $k$-th particle. Writing the average explicitly and summing over all spatial directions one gets
\begin{multline}
\bar{\mathbf{J}}^{k,j} 
=\\=
\frac{m_0}{\gamma M}G m_k m_j
\int \dd[3]{\bz_j}\dd[3]{\bz_k}\dd[3]{\by} \sigma^{(j,k)}_{t_c^-} (\bz_j,\bz_k;\bz_j,\bz_k) g_{r_C} (\by+\bz_k-\bz_j) \frac{\by}{\abs{\by}^2}
\prtq{\frac{1}{\abs{\by}}\erf \prt{\frac{\abs{\by}}{r_C\sqrt{2}}}
	-4 \pi r_C^2 g_{r_C} (\by)},
\end{multline}
where $\sigma^{(j,k)}_{t_c^-} (\bz_j,\bz_k;\bz_j',\bz_k')$ is the reduced density matrix of the particles $j$ and $k$ in position representation and we also made the change of variables $\by=\bx-\bz_k$. We decided to use $\sigma$ instead of $\rho$ to denote the density matrix at time $t_c^-$ to distinguish it from the density matrix of the master equation, which is obtained by averaging over the noise.
Using the spherical symmetry of $g_{r_C} (\by)$, a simple change of variable $\by \rightarrow -\by$ shows that $\bar{\mathbf{J}}^{k,j} = -\bar{\mathbf{J}}^{j,k}$. Therefore, at the average (over state and noise) level, pair of impulses neutralizes and it follows that the average total momentum is conserved.

Let us now derive the average force between two particles. When considering $N$ particles whose total mass amounts to $M$, the probability for a collapse happening anywhere in a time $\dd{t}$ is given by [cf. Eq.~\eqref{eq:PoissonProcessCharacterization}]:
\begin{equation}
\mbE\prtq{ \int \dd{N}_t (\bx) \dd[3]{\bx}} = \int \dd[3]{\bx} \frac{\gamma}{m_0}\ev{\hmu_{r_C} (\bx)}{\psi_t} \dd{t}
=
\frac{\gamma M}{m_0} \dd{t},
\end{equation}
which means that we can obtain the average force exerted by particle $j$ on particle $k$ by multiplying the average impulse by the rate $\gamma M/m_0$, obtaining
\begin{equation}\label{APPeq:AverageForceTrajectoryLevel}
\bar{\mathbf{F}}^{k,j} 
=
G m_k m_j
\int \dd[3]{\bz_j}\dd[3]{\bz_k}\dd[3]{\by} \sigma^{(j,k)}_{t_c^-} (\bz_j,\bz_k;\bz_j,\bz_k) g_{r_C} (\by+\bz_k-\bz_j) \frac{\by}{\abs{\by}^2}
\prtq{\frac{1}{\abs{\by}}\erf \prt{\frac{\abs{\by}}{r_C\sqrt{2}}}
	- 4 \pi r_C^2 g_{r_C} (\by)}.
\end{equation}
Finally, when dealing with the average dynamics (without following a specific trajectory of the system), one can see that $\sigma^{(j,k)}_{t_c^-} (\bz_j,\bz_k;\bz_j,\bz_k)= \mbE [\sigma^{(j,k)}_{t_c^+} (\bz_j,\bz_k;\bz_j,\bz_k)]$. This follows from the fact that $\sigma^{(j,k)}_{t} (\bz_j,\bz_k;\bz_j,\bz_k)$ can be seen as the joint probability distribution of the observables $\hbq_j$ and $\hbq_k$, which are conserved according to Eq.~\eqref{eq:PSLDynamicsWithGravityParticles}. In Eq.~\eqref{eq:PSLDynamicsWithGravityParticles}, the standard Hamiltonian is neglected, but this has no consequence for the equality $\sigma^{(j,k)}_{t_c^-} (\bz_j,\bz_k;\bz_j,\bz_k)= \mbE [\sigma^{(j,k)}_{t_c^+} (\bz_j,\bz_k;\bz_j,\bz_k)]$, because the standard Hamiltonian cannot have dynamical effects in an infinitesimal time. This reasoning allows us to write $\rho^{(j,k)}_{t} (\bz_j,\bz_k;\bz_j,\bz_k)$ in place of $\sigma^{(j,k)}_{t_c^-} (\bz_j,\bz_k;\bz_j,\bz_k)$, thus giving Eq.~\eqref{eq:PairwiseNewtonianForceGPSL} of the main text.

\begin{figure}[t]
\centering
\includegraphics[width=0.8\textwidth]{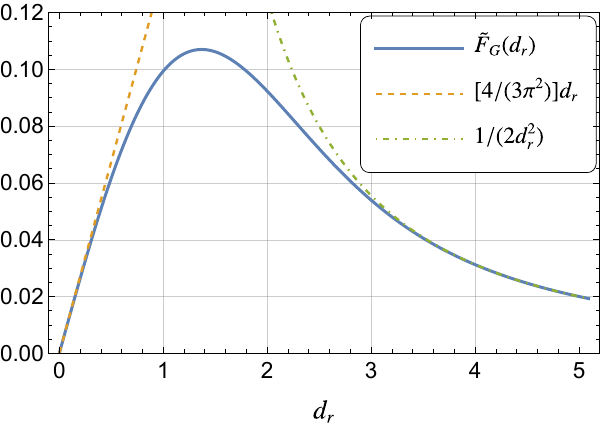}
\caption{The plot shows the function $\tl{F}_G (d_r)$, defined in Eq.~\eqref{APPeq:NumericalAverageForces} and its perturbative forms defined in Eq.~\eqref{APPeq:ExtremeRegimesAverageForces}.}
\label{fig:GravForceAverage}
\end{figure}

For each couple of points $\bz_k$ and $\bz_j$, the integral appearing in Eq.~\eqref{APPeq:AverageForceTrajectoryLevel} gives rise to a vector quantity that is directed from $\bz_k$ to $\bz_j$. This can be seen by assuming that $\bd_{j,k} := \bz_j-\bz_k$ lies, say, on the $y_3$ axis. Then, the integration for the two orthogonal components (to $\hat{e}_3$) gives zero because, for each value of $y_3$, all functions appearing in the integral are spherically symmetric in the $y_1$-$y_2$ plane except for $y_j$ with $j=1,2$, which is an odd function. Thus, the integral vanishes unless $j=3$, that is when we are computing the force along the direction parallel to $\bz_j-\bz_k$. The fact that the force is always directed from $\bz_k$ to $\bz_j$ can be seen from numerical integration. In fact, the integral may be rewritten as follows by using spherical coordinates and integrating the angular parts:
\begin{multline}\label{APPeq:NumericalAverageForces}
\int \dd[3]{\bx} g_{r_C} (\by-\bd_{j,k}) \frac{\by}{\abs{\by}^2}
\prtq{\frac{1}{\abs{\by}}\erf \prt{\frac{\abs{\by}}{r_C\sqrt{2}}}
	- 4 \pi r_C^2 g_{r_C} (\by)}
=\\=
\frac{e^{-d_r^2}}{4 \pi d_r^2 r_C^2}\int_0^{\infty} \dd{r} \frac{e^{-r^2}}{r}\prtq{e^{-2 d_r r}\prt{1+2 d_r r} - e^{+2 d_r r}\prt{1 - 2 d_r r}}\prtq{\frac{\sqrt{\pi}}{r}\erf (r) -2 e^{-r^2}}
=: \frac{\tl{F}_G (d_r)}{r_C^2},
\end{multline}
where $d_r := \abs{\bd_{j,k}}/(r_C \sqrt{2})$ so that everything is written in terms of a-dimensional quantities except for the prefactor $1/r_C^2$. The quantity $\tl{F}_G (d_r)$ is plotted in Fig.~\ref{fig:GravForceAverage}, where we also show that
\begin{equation}\label{APPeq:ExtremeRegimesAverageForces}
\tl{F}_G (d_r \ll 1) \simeq \frac{4}{3 \pi^2} d_r,
\qquad
\tl{F}_G (d_r \gg 1) \simeq \frac{\sign(d_r)}{2}\frac{1}{d_r^2}.
\end{equation}
Therefore, when considering two particles such that $\rho^{(j,k)}_{t} (\bz_j,\bz_k;\bz_j,\bz_k)$ is non-negligible only when $\abs{\bz_j-\bz_k} \gg r_C$ one correctly gets that
\begin{equation}
\bar{\mathbf{F}}^{k,j} 
\simeq
\int \dd[3]{\bz_j}\dd[3]{\bz_k} \rho^{(j,k)}_{t} (\bz_j,\bz_k;\bz_j,\bz_k) \prtq{G m_k m_j \frac{\bz_j-\bz_k}{\abs{\bz_j-\bz_k}^3}}.
\end{equation}

\clearpage
\section{Reduced Dynamics of a Single Particle Distant from other Masses\label{APPsec:ReducedParticleDynamics}}

In this appendix, we derive Eq.~\eqref{eq:SingleParticleGravitationalUnitaryReducedDynamics} of the main text. We want to check that a particle far enough from all other masses behaves as expected according to the unitary-like term of Eq.~\eqref{eq:PerturbativeGPSL_MasterEquation}. 
To this end, with a certain abuse of notation, we write the smeared mass operator as $m g_{r_C}(\bx-\hbq) + \hmu_{r_C} (\bx)$, the non-smeared version in a similar way, and the square root of the smeared mass operator as $\sqrt{m g_{r_C}(\bx-\hbq)} + \sqrt{\hmu_{r_C} (\bx)}$. This writing of the square root is possible because the particle is assumed to be much more distant than $r_C$ from all other masses. For the same reason we take the state to be $\rho_t = \rho^{(p)}_t \otimes \rho^{(\mu)}_t$ where $\rho^{(p)}_t$ is the density matrix of the particle and $\rho^{(\mu)}_t$ that of the other masses\footnote{A more general case where it makes sense to say that the particle is much more distant than $r_C$ from the rest of the masses is if $\rho_t \sum_k a_k \rho^{(p)}_{t,k} \otimes \rho^{(\mu)}_{t,k}$ where this distance is realized for every $k$.}. We can also separate the potential as $\hV (\bx) = V_p (\bx) + V_\mu (\bx)$ with $V_p (\bx)= - G m_0 m \abs{\bx-\hbq}^{-1}\erf \prtq{\abs{\bx-\hbq}/(\sqrt{2}r_C)}$, and we have that for every $\bx$ one has $\sqrt{g_{r_C}(\bx-\hbq)}\rho_t \sqrt{\hmu_{r_C} (\bx)}\simeq 0$. We then have
\begin{multline}\label{APPeq:PSLGravityTestParticle}
-\frac{i}{\hbar}\frac{1}{m_0}\int \dd[3]{\bx} \Tr_{\mu} \prtg{\comm{\hV (\bx)}{\sqrt{g_{r_C} (\bx-\hbq)}\rho_t\sqrt{g_{r_C} (\bx-\hbq)}+\sqrt{\hmu_{r_C} (\bx)}\rho_t\sqrt{\hmu_{r_C} (\bx)}}}
=\\=
-\frac{i}{\hbar}\frac{1}{m_0}\int \dd[3]{\bx} \comm{V_p (\bx)}{\sqrt{g_{r_C}(\bx-\hbq)}\rho^{(p)}_t \sqrt{g_{r_C}(\bx-\hbq)}}
-\frac{i}{\hbar}\frac{1}{m_0}\int \dd[3]{\bx} \comm{V_p (\bx) \Tr \{ \hmu_{r_C} (\bx) \rho_t^{(\mu)}\}}{\rho^{(p)}_t}.
\end{multline}

The first term of Eq.~\eqref{APPeq:PSLGravityTestParticle} seems a self-interaction term which however gives a null contribution to the dynamics. This can be proven by writing the integral as follows
\begin{multline}\label{eq:NoSelfInteractionSingleParticle}
\mel{\bx}{\int \dd[3]{\bz} \comm{V_p (\bx)}{\sqrt{g_{r_C}(\bz-\hbq)}\rho^{(p)}_t \sqrt{g_{r_C}(\bz-\hbq)}}}{\by}
\propto\\ \propto
\rho^{(p)}_t(\bx,\by)\int \dd[3]{\bz} \sqrt{g_{r_C} (\bx-\bz)g_{r_C} (\by-\bz)}\prt{\frac{\erf \prt{\abs{\bx-\bz}/(\sqrt{2}r_C)}}{\abs{\bx-\bz}}-\frac{\erf \prt{\abs{\by-\bz}/(\sqrt{2}r_C)}}{\abs{\by-\bz}}}
\propto\\ \propto
\rho^{(p)}_t(\bx,\by)\int \dd[3]{\bz} \sqrt{g_{r_C} (\bz)g_{r_C} (\bz+\bd)}\prt{\frac{\erf \prt{\abs{\bz}/(\sqrt{2}r_C)}}{\abs{\bz}}-\frac{\erf \prt{\abs{\bz+\bd}/(\sqrt{2}r_C)}}{\abs{\bz+\bd}}}
\propto\\ \propto
\rho^{(p)}_t(\bx,\by)\prtq{\int \dd[3]{\bz} \sqrt{g_{r_C} (\bz)g_{r_C} (\bz+\bd)}\frac{\erf \prt{\abs{\bz}/(\sqrt{2}r_C)}}{\abs{\bz}}
	-\int \dd[3]{\bz} \sqrt{g_{r_C} (\bz)g_{r_C} (\bz-\bd)}\frac{\erf \prt{\abs{\bz}/(\sqrt{2}r_C)}}{\abs{\bz}}}=0.
\end{multline}
In the last line, the two integrals give the same result because it depends only on $\abs{\bd}=\abs{\bx-\by}$ due to the spherical symmetry of $g_{r_C} (\bx)$ and $\erf(\abs{\bx})$.

The second term of Eq.~\eqref{APPeq:PSLGravityTestParticle} represents the effect of the other masses on our test particle. Since the distance is much higher than $r_C$ we can make the substitutions $g_{r_C}(\bx)\rightarrow \delta (\bx)$ inside $V_p (\bx)$, and $\Tr \{ \hmu_{r_C} (\bx) \rho_t^{(\mu)}\} \rightarrow \Tr \{ \hmu (\bx) \rho_t^{(\mu)}\}$. So, we get
\begin{equation}
-\frac{i}{\hbar}\frac{1}{m_0}\int \dd[3]{\bx} \comm{V_p (\bx)\Tr \{ \hmu_{r_C} (\bx) \rho_t^{(\mu)}\}}{\rho^{(p)}_t}
\simeq
-\frac{i}{\hbar}\comm{\int \dd[3]{\bx}  \frac{-G m \Tr \{ \hmu (\bx) \rho_t^{(\mu)}\}}{\abs{\bx-\hbq}}}{\rho^{(p)}_t},
\end{equation}
which gives Eq.~\eqref{eq:SingleParticleGravitationalUnitaryReducedDynamics} of the main text.

\clearpage
\section{Single Particle Decoherence in GPSL\label{APPsec:SingleParticleDecoherencePSL}}

In this appendix, we compute the decoherence rate appearing in Eq.~\eqref{eq:SingleParticlePSLGravitationalDecoherence_Evaluated} of the main text. 
We start from
\begin{equation}
\Gamma (\bx,\by) = \frac{\gamma m}{m_0}
\prtq{1-\int \dd[3]{\bz} e^{i r_p [f(\bx-\bz)-f(\by-\bz)]}\sqrt{g_{r_C}(\bx-\bz)g_{r_C}(\by-\bz)}},
\end{equation}
which, being real, can be written as
\begin{equation}
\Gamma (\bx,\by) = \frac{\gamma m}{m_0}
\prtq{1-\int \dd[3]{\bz} \cos(r_p [f(\bx-\bz)-f(\by-\bz)])\sqrt{g_{r_C}(\bx-\bz)g_{r_C}(\by-\bz)}}.
\end{equation}
Expanding the cosine to second order we get
\begin{equation}
\Gamma (\bx,\by) = \frac{\gamma m}{m_0}
\prtq{1-\int \dd[3]{\bz} \prt{1 - \frac{r_p^2}{2} \prtq{  f(\bx-\bz)-f(\by-\bz)}}\sqrt{g_{r_C}(\bx-\bz)g_{r_C}(\by-\bz)}}.
\end{equation}
The zero-order part gives the collapse contribution to the decoherence rate and is computed in Ref.~\cite{Piccione2025ExploringMassDependence}. Here we compute the integral appearing in the second order term:
\begin{equation}\label{APPeq:IntegralToComputeSingleParticle}
\int \dd[3]{\bz} \sqrt{g_{r_C} (\bx-\bz)g_{r_C} (\by-\bz)}\prt{\frac{\erf \prt{\abs{\bx-\bz}/(\sqrt{2}r_C)}}{\abs{\bx-\bz}}-\frac{\erf \prt{\abs{\by-\bz}/(\sqrt{2}r_C)}}{\abs{\by-\bz}}}^2.
\end{equation}

First we recall that
\begin{equation}
\int \dd[3]{\bz'} \frac{g_{r_C} (\bx-\bz')}{\abs{\bz-\bz'}} = \frac{1}{\abs{\bx-\bz}}\erf \prt{\abs{\bx-\bz}/(\sqrt{2}r_C)},
\end{equation}
which can be used to put Eq.~\eqref{APPeq:IntegralToComputeSingleParticle} in a more useful form.
As a result of the fact the evaluation of Eq.~\eqref{APPeq:IntegralToComputeSingleParticle} can only depend on $d=\abs{\bx-\by}$, we write it in the following way
\begin{equation}
\int \dd[3]{\bz}\dd[3]{\bz'}\dd[3]{\bz''} G_1(\bz,d)G_2(\bz',d)G_2(\bz'',d)\frac{1}{\abs{\bz-\bz'}}\frac{1}{\abs{\bz-\bz''}},
\end{equation}
where $G_1 (\bz,d)= \sqrt{g_{r_C} (\bz)g_{r_C} (\bz-d\hat{e}_3)}$, and $G_2(\bz',d) = \prtq{g_{r_C} (\bz')-g_{r_C} (\bz'-d\hat{e}_3)}$. This integral can then be rewritten as follows by remembering that $\mcF \prt{\abs{\bz-\bz'}^{-1}} (\bk,\bk') = 4\pi \bk^{-2} \delta (\bk+\bk')$:
\begin{equation}
\int \dd[3]{\bz}\dd[3]{\bz'}\dd[3]{\bz''} G_1(\bz,d)G_2(\bz',d)G_2(\bz'',d)\frac{1}{\abs{\bz-\bz'}}\frac{1}{\abs{\bz-\bz''}}
=
\frac{16 \pi^2}{(2\pi)^{3/2}} \int \dd[3]{\bk}\dd[3]{\bv} 
\frac{\tl{G}_1 (-(\bk+\bv),d)\tl{G}_2 (\bk,d)\tl{G}_2 (\bv,d)}{\bk^2 \bv^2}.
\end{equation}
Now we compute the Fourier transform of each of these functions:
\begin{equation}
\tl{G}_1 (\bk) = e^{-d^2/(8 r_C^2)}\frac{e^{i d k_3/2}}{(2\pi)^{3/2}}e^{-\bk^2 r_C^2/2},
\qquad
\tl{G}_2 (\bk) = \frac{1-e^{i d k_3}}{(2\pi)^{3/2}}e^{-\bk^2 r_C^2/2}.
\end{equation}
So, the integral can be written as
\begin{multline}
-\frac{64 \pi^2}{(2\pi)^{6}}e^{-d^2/8 r_C^2}
\int \dd[3]{\bk}\dd[3]{\bv} 
\sin(\frac{d k_3}{2})\sin(\frac{d v_3}{2})\frac{e^{-r_C^2(\bk^2+\bv^2+\bk\cdot\bv)}}{\bk^2 \bv^2}.
=\\=
\frac{64 \pi^2}{(2\pi)^{6}}e^{-d^2/8 r_C^2}
\int \dd[3]{\bk}\dd[3]{\bv} 
\sin(\frac{d k_3}{2})\sin(\frac{d v_3}{2})\frac{e^{-r_C^2(\bk^2+\bv^2-\bk\cdot\bv)}}{\bk^2 \bv^2}.
=\\=
\frac{1}{\pi^4}\frac{e^{-\td^2/2}}{r_C^2}
\int \dd[3]{\tl{\bk}}\dd[3]{\tl{\bv}} 
\sin(\td \tl{k}_3)\sin(\td \tl{v}_3)\frac{e^{-(\tl{\bk}^2+\tl{\bv}^2-\tl{\bk}\cdot\tl{\bv})}}{\tl{\bk}^2 \tl{\bv}^2}
=
\frac{e^{-\td^2/2}}{\pi^4 r_C^2} \tl{F}(\td),
\end{multline}
where in the last line we used $\td =d/(2 r_C)$, $\tl{\bk}=r_C \bk$, and $\tl{\bv} = r_C \bv$.

It does not seem possible to evaluate $\tl{F} (\td)$ analytically. However, we can compute its expansion for $\td \ll 1$. This is so because by the time $\sin(\td \tl{k_3})$ is not anymore well approximated by its Taylor expansion, the exponential is such that we are in the region of very small contributions to the integral. Numerically, we get $\tl{F} (\td)\vert_{\td \ll 1} \simeq 4.49\times \td^2$. On the other hand, we know that $\lim_{\td \rightarrow \infty} F(\td) = 0$ because the integrand becomes rapidly oscillating. By numerically plotting the function and then fitting it, we discover that it decays exponentially fast to zero (see the caption of Fig.~\ref{fig:plotFtilde}). The numerical integration is more easily done by writing $\tl{F} (\td)$ as follows:
\begin{multline}
F(d) = 4 \pi^2 \int_0^\pi \dd{\theta_k}\dd{\theta_v} \int_0^{+\infty} \dd{k}\dd{v} e^{-(k^2 + v^2 - kv \cos(\theta_k)\cos(\theta_v))} 
\times \\ \times
I_0 (kv \sin(\theta_k)\sin(\theta_v))
\sin(\theta_k)\sin(\theta_v)
\sin(d k \cos(\theta_k))\sin(d v \cos(\theta_v)),
\end{multline}
where $I_n (x)$ is the modified Bessel function of the first kind\footnote{See \url{https://reference.wolfram.com/language/ref/BesselI.html.en}}. Moreover, we used the \enquote{Adaptive MonteCarlo} built-in method in Mathematica to produce the plot of Fig.~\ref{fig:plotFtilde} because standard numerical integration methods were not converging.

\clearpage
\section{Decoherence of a Spherical Rigid Body\label{APPsec:DecoherenceRigidSphere}}

In this appendix, we show how to compute the decoherence rates appearing in Sec.~\ref{Sec:RigidSphere} of the main text. Using Eq.~\eqref{eq:RigidBodyMassDensityOperator}, the decoherence due to gravity for GPSL and for the continuous weak monitoring models based on CSL and DP is given by [cf. Eq.~\eqref{eq:PerturbativeGPSL_MasterEquation} for the GPSL model and Eq.~\eqref{APPeq:GravitationalMasterEquationTDModels} for the continuous weak monitoring ones]
\begin{equations}\label{APPeq:DecoherenceRatesRigidBody}
\prtq{\dv{t} \rho_t (\bX,\bY)}_{\rm Grav} &= - \Gamma_{G} (\bD) \rho_{t} (\bX,\bY),\\
\Gamma_{G}^{\rm DP} (\bD) &\equiv 
\frac{G}{2\hbar}\int \dd[3]{\bz}\varrho_{r_C} (\bz) \int \dd[3]{\bz'} \frac{1}{\abs{\bz-\bz'}}\prtq{\varrho_{r_C}(\bz') - \varrho_{r_C}(\bz'+\bD)},\\
\Gamma_{G}^{\rm GPSL} (\bD) &\equiv 
\frac{G^2 m_0}{2\hbar^2 \gamma_{\rm PSL}} \int \dd[3]{\bz} \sqrt{\varrho_{r_C} (\bz)\varrho_{r_C} (\bz+\bD)} 
\prtq{\int \dd[3]{\bz'}\frac{1}{\abs{\bz-\bz'}}\prtq{\varrho_{r_C} (\bz') - \varrho_{r_C} (\bz'+\bD)}}^2,\\
\Gamma_{G}^{\rm CSL} (\bD) &\equiv 
\frac{m_0^2 G^2}{8 \hbar^2 \gamma_{\rm CSL}} \int \dd[3]{\bz} \prtq{\int \dd[3]{\bz'}\frac{1}{\abs{\bz-\bz'}}\prtq{\varrho_{r_C} (\bz') - \varrho_{r_C} (\bz'+\bD)}}^2,\\
\end{equations}
where we introduced $\bD \equiv \bX-\bY$. In the following, we will make the approximation $\varrho_{r_C} (\bx) \simeq (M/V) \chi_V (\bx)$, where $M$ is the mass of the rigid body, $V$ is its volume, and $\chi_V (\bx)$ is the indicator function associated to the volume of space occupied by the rigid body around its center of mass.

To compute the decoherence rate of a spherical rigid body we first recall how to generally write the Fourier transform of a spherically symmetric function:
\begin{equation}
\tl{f}(\bk) 
= \frac{1}{(2\pi)^{3/2}}\int \dd[3]{\bx} f(\abs{\bx})e^{i \bk\cdot \bx}
= \frac{2\pi}{(2\pi)^{3/2}}\int_0^{\infty} \dd{r} r^2 f(r) \int_0^{\pi} \dd{\theta} \sin(\theta)e^{i k r \cos(\theta)}
= \frac{4 \pi}{(2\pi)^{3/2}}\int_0^{\infty} \dd{r} r^2 \frac{\sin(k r)}{k r} f(r).
\end{equation}
Using this, we can compute the Fourier transform of the indicator function of a sphere of radius $R$:
\begin{equation}\label{APPeq:RigidSphereIndicatorFunctionFourierTransform}
\tl{\chi}_R (k) 
= \frac{4\pi}{(2\pi)^{3/2}}\int_0^{R} \dd{r} r^2 \frac{\sin(k r)}{k r}
= \frac{\sqrt{2/\pi}}{k^3}\prtq{\sin(k R) - k R \cos(k R)}.
\end{equation}

To compute the decoherence rate for all models we need to compute the following integral
\begin{multline}
\int \dd[3]{\bz'} \frac{\chi_R (\bz')}{\abs{\bz-\bz'}}
=
\int \dd[3]{\bk} \frac{4 \pi}{(2\pi)^{3/2}\bk^2} \tl{\chi}_R (\abs{\bk}) e^{i \bk \cdot \bz}
=\\=
\frac{16 \pi^2}{(2\pi)^{3/2}} \int_0^{\infty} \dd{k} \frac{\sin(k r_{\bz})}{k r_{\bz}}\tl{\chi}_R (k)
=
\frac{16 \pi^2}{(2\pi)^{3/2}} R^2 \int_0^{\infty} \dd{\tl{k}} \frac{\sin(\tl{k} (r_{\bz}/R))}{\tl{k} (r_{\bz}/R)} \tl{\chi}_1 (\tl{k})
=
(2 \pi R)^2 F_{\rm Sp} (r_{\bz}/R).
\end{multline}
where 
\begin{equation}
F_{\rm Sp} (x) = \frac{1}{\pi}\prtq{\Theta (1-x)\frac{3-x^2}{6} + \Theta(x-1)\frac{1}{3x}}.
\end{equation}
Similarly, we can compute
\begin{equation}
\int \dd[3]{\bz'} \frac{\chi_R (\bz'+\bD)}{\abs{\bz-\bz'}}
=
\int \dd[3]{\by} \frac{\chi_R (\by)}{\abs{(\bz+\bD)-\by}}
=
(2 \pi R)^2 F_{\rm Sp} (r_{\bz+\bD}/R),
\end{equation}
where $r_{\bz+\bD}$ is the radial component of the vector $\bz+\bD$.

For the DP model, we have to compute
\begin{equation}
\int \dd[3]{\bz} \chi_R (\bz) F_{\rm Sp} \prt{\frac{r_{\bz}}{R}}
=
4 \pi \int_0^{R} r^2 F_{\rm Sp} \prt{\frac{r_{\bz}}{R}}
=\frac{2}{15}(2\pi)^{3/2} R^3,
\end{equation}
and
\begin{multline}
\int \dd[3]{\bz} \chi_R (\bz) F_{\rm Sp} \prt{\frac{r_{\bz+\bD}}{R}}
=
\int \dd[3]{\by} \chi_R (\by-\bD) F_{\rm Sp} \prt{\frac{r_{\by}}{R}}
=
R^3 \int \dd[3]{\bk} \tl{\chi}_R (\abs{\bk}) e^{i\bk\cdot \bD} \tl{F}_{\rm Sp} (\abs{\bk} R)
=\\=
4 \pi R^3 \int_0^{\infty} \dd{k} k^2 \frac{\sin(k r_{\bD})}{k r_{\bD}} \tl{\chi}_R (k) \tl{F}_{\rm Sp} (k R)
=
4 \pi R^3 \int_0^{\infty} \dd{\tl{k}} \tl{k}^2 \frac{\sin(\tl{k} (r_{\bD}/R))}{\tl{k} (r_{\bD}/R)} \tl{\chi}_1 (\tl{k}) \tl{F}_{\rm Sp} (\tl{k})
=\\=
\frac{(2\pi)^{3/2}}{9}R^3\prtq{\Theta\prt{2-\frac{r_{\bD}}{R}}\frac{192-(r_{\bD}/R)^2(80-30(r_{\bD}/R) + (r_{\bD}/R)^3)}{160} + \Theta\prt{\frac{r_{\bD}}{R}-2}\frac{1}{r_{\bD}/R}},
\end{multline}
where $\tl{F}_{\rm Sp} (\tl{k})= k^{-5}\prtq{\sin(k)-k\cos(k)}$. Substituting back into Eq.~\eqref{APPeq:DecoherenceRatesRigidBody} with $\varrho (\bx) = (3/4) M \pi^{-1} R^{-3} \chi_R (\bx)$ and making some algebraic simplifications one gets half of the decoherence rate reported in Eq.~\eqref{eq:RigidSphereDecoherenceTDDP}. We recall that in the continuous weak monitoring model based on the DP model the collapse decoherence rate is exactly equal to the gravitational one.

To compute the decoherence in the GPSL model we need the following integral [cf. Eq.~\eqref{APPeq:DecoherenceRatesRigidBody}]
\begin{equation}\label{APPeq:GPSLCalculationStepToUseInCSL}
\int \dd[3]{\bz} \chi_R (\bz) \chi_{R} (\bz+\bD)\prtq{F_{\rm Sp} \prt{\frac{r_{\bz}}{R}}-F_{\rm Sp} \prt{\frac{r_{\bz+\bD}}{R}}}^2
=
2\int \dd[3]{\bz} \chi_R (\bz) \chi_{R} (\bz+\bD)F_{\rm Sp} \prt{\frac{r_{\bz}}{R}}\prtq{F_{\rm Sp} \prt{\frac{r_{\bz}}{R}}-F_{\rm Sp} \prt{\frac{r_{\bz+\bD}}{R}}}.
\end{equation}
Since both $\bz$ and $\bz+\bD$ have to be within a radius $R$ of the origin we can take the function $F_{\rm Sp} (\bx)$ as $F_{\rm Sp} (\bx) = (3-\bx^2)/(6\pi)$. Therefore, introducing $\tl{\bz}=\bz/R$ and $\tl{\bD}=\bD/R$ we can write the integral in two equivalent forms
\begin{equations}
\int \dd[3]{\bz} \chi_R (\bz) \chi_{R} (\bz+\bD)\prtq{F_{\rm Sp} \prt{\frac{r_{\bz}}{R}}-F_{\rm Sp} \prt{\frac{r_{\bz+\bD}}{R}}}^2
&=
\frac{R^3}{18 \pi^2}\int \dd[3]{\tl{\bz}} \chi_1 (\tl{\bz}) \chi_{1} (\tl{\bz}+\tl{\bD}) \prt{3 - \tl{\bz}^2}\prtq{\tl{\bD}^2 - 2 \tl{\bz}\cdot \tl{\bD}},\\
&=
\frac{R^3}{18 \pi^2}\int \dd[3]{\tl{\bz}} \chi_1 (\tl{\bz}) \chi_{1} (\tl{\bz}+\tl{\bD}) \prtq{\tl{\bD}^2 - 2 \tl{\bz}\cdot \tl{\bD}}^2.
\end{equations}
Let us now choose $\tl{\bD} = 2 d \hat{e_3}$ with $0 \leq d \leq 1$. The integral then becomes
\begin{multline}
R^3 \frac{16 d^2}{18 \pi^2}\int \dd[3]{\tl{\bz}} \chi_1 (\tl{\bz}) \chi_{1} (\tl{\bz}+\tl{\bD}) \prtq{d-z_3}^2
=
R^3 \frac{16 d^2}{9 \pi} \int_{0}^{\sqrt{1-d^2}} \dd{\rho} \rho \int_{-\sqrt{1-\rho^2}}^{\sqrt{1-\rho^2}- 2 d} \dd{z_3} \prtq{d-z_3}^2
=\\=
R^3 \frac{4 d^2}{27 \pi}\prtq{\prt{3+60 d^2}\arccos(d) - d\sqrt{1-d^2}(13+50 d^2)}\Theta(1-d).
\end{multline}
Substituting back into Eq.~\eqref{APPeq:DecoherenceRatesRigidBody} with $\varrho (\bx) = [(4/3)\pi R^{3}]^{-1} M \chi_R (\bx)$ one gets the gravitational part of the decoherence rate reported in Eq.~\eqref{eq:RigidSphereDecoherenceGPSL}. The collapse part of the decoherence rate is given by
\begin{equation}
\Gamma^{\rm CM}_{\rm Collapse} (\bD)
=
\gamma \frac{M}{m_0}\prtq{1-\frac{1}{V}\int\dd[3]{\bx} \chi_R (\bx)\chi_R (\bx+\bD)}
=
\gamma \frac{M}{m_0}\prtg{1-\Theta(1-D)\prtq{1-\frac{3}{2}D+ \frac{1}{2}D^3}}.
\end{equation}
where $D=\abs{D}/(2R)$. Summing the two parts one gets Eq.~\eqref{eq:RigidSphereDecoherenceGPSL}.

Finally, to compute the gravitational decoherence in the CSL model, we have to do the same calculation steps as in Appendix~\ref{APPSubsec:TDModelSingleParticlesDecoherenceRate}. In particular, one gets
\begin{multline}
\Gamma_{G}^{\rm CSL} (\bD) =
\frac{m_0^2 M^2 G^2}{8 \hbar^2 \gamma_{\rm CSL} V^2} \int \dd[3]{\bz}\dd[3]{\bz'}\dd[3]{\bz''}\frac{1}{\abs{\bz-\bz'}}\frac{1}{\abs{\bz-\bz''}}\prtq{\chi_V (\bz')-\chi_V (\bz'-D\hat{e}_3)}\prtq{\chi_V (\bz'')-\chi_V (\bz'-D\hat{e}_3)}
=\\=
\frac{2 \pi^2 m_0^2 M^2 G^2}{\hbar^2 \gamma_{\rm CSL} V^2}
\int \dd[3]{\bk} 
\frac{\tl{G} (\bk,d)\tl{G} (-\bk,d)}{\bk^4},
\end{multline}
where
\begin{equation}
\tl{G} (\bk,d) = \tl{\chi}_R (k)[1-e^{i D k_3}]
\implies
\tl{G} (\bk,d)\tl{G} (-\bk,d) = 2\tl{\chi}_R^2(k)[1-\cos(D k_3)].
\end{equation}
Then [cf. Eq.~\eqref{APPeq:RigidSphereIndicatorFunctionFourierTransform}],
\begin{multline}
\Gamma_{G}^{\rm CSL} (\bD)
= \frac{4 \pi^2 m_0^2 M^2 G^2}{\hbar^2 \gamma_{\rm CSL} V^2}
\int \dd[3]{\bk} 
\frac{\tl{\chi}_R^2(k)[1-\cos(D k_3)]}{\bk^4}
=\\= 
\frac{4 \pi^2 m_0^2 M^2 G^2}{\hbar^2 \gamma_{\rm CSL} V^2} \times \frac{2 \pi R^7}{315}\prtq{\Theta(1-D)\prt{56 D^2-28 D^4+14D^5 - D^7} + \Theta(D-1)\prt{70 D - 36 + \frac{7}{D}}}.
\end{multline}
The collapse part of the decoherence rate for the CSL model can be computed as
\begin{equation}
\Gamma^{\rm CM}_{\rm CSL} (\bD)
=
\frac{\gamma_{\rm CSL}}{V} \frac{M^2}{m_0^2}\prtq{1-\frac{1}{V}\int\dd[3]{\bx} \chi_R (\bx)\chi_R (\bx+\bD)}
=
\frac{3\gamma_{\rm CSL}}{4 \pi R^3}\frac{M^2}{m_0^2}\prtg{1-\Theta(1-D)\prtq{1-\frac{3}{2}D+ \frac{1}{2}D^3}}.
\end{equation}

\end{document}